%% file: two_parameter_entropies.tex
\documentclass[a4paper,12pt]{article}
\usepackage{amssymb,amsmath}
\usepackage{graphics}
\def\beq{\begin{equation}}
\def\eeq{\end{equation}}
\def\bea{\begin{eqnarray}}
\def\eea{\end{eqnarray}}
\def\nn{\nonumber}
\makeatletter
  \def\@cite#1#2{${\mbox{#1\if@tempswa , #2\fi}}$}
\makeatother

\makeatletter
  \def\@biblabel#1{$^{\mbox{#1}}$}
\makeatother
\begin{document}
%
%
%
%
\thispagestyle{empty}
\vspace*{3cm}
\begin{center}
{\LARGE\sf Adiabatic thermostatistics of the two parameter entropy and the role of 
Lambert's $W$-function in its applications} \\

\bigskip\bigskip
R. Chandrashekar${}^{\dagger}$ and J. Segar${}^{\ddagger}$

\bigskip
\textit{
${}^{\dagger}$The Institute of Mathematical Sciences, \\
C.I.T Campus, Taramani, \\
Chennai 600 113, India \\}
\bigskip
\textit{
${}^{\ddagger}$Ramakrishna Mission Vivekananda College\\
Mylapore\\
Chennai 600 004, India.
}
\end{center}

\vfill
\begin{abstract}
A unified framework to describe the adiabatic class of ensembles in the generalized statistical mechanics
based on Schw\"{a}mmle-Tsallis two parameter $(q,q^{\prime})$ entropy is proposed.  The generalized form
of the equipartition theorem, virial theorem and the adiabatic theorem are derived.  Each member of the 
class of ensembles is illustrated using the classical nonrelativistic ideal gas and we observe that the
heat functions could be written in terms of the Lambert's $W$-function in the large $N$ limit. 
In the microcanonical ensemble we study the effect of gravitational field on classical nonrelativistic 
ideal gas and a system of hard rods in one dimension and compute their respective internal energy and 
specific heat. We found that the specific heat can take both positive and negative values depending on 
the range of the deformation parameters, unlike the case of one parameter Tsallis entropy.  
\end{abstract}

PACS Number(s): 05.20.Gg, 05.70.Ce, 02.30.Gp 
\newpage
\setcounter{page}{1}
%
%
%
\setcounter{equation}{0}
\section{Introduction}
\label{Intro}
A system in thermodynamic equilibrium with its surroundings can be described using three macroscopic variables 
corresponding to the thermal, mechanical, and the chemical equilibrium.  For each fixed value of these macroscopic
variables (macrostate) we have many possible microscopic configurations (microstates).  
A collection of systems existing in the various possible microstates, but characterized by the same macroscopic 
variables is called an ensemble. The thermal, mechanical, and the chemical parameters can be choosen 
between either an extensive variable or an intensive variable and so, we have, in total eight different ensembles. 
The eight ensembles are further divided into two classes namely the isothermal class for which the thermal 
equilibrium variable is the temperature and the adiabatic class for which the thermal equilibrium variable is the 
heat function. The isothermal class comprises of the the canonical $(N,V,T)$, isothermal-isobaric $(N,P,T)$, 
grandcanonical $(\mu,V,T)$ and the generalized ensemble $(\mu,P,T)$. All the individual members of the adiabatic 
class of ensembles have same value for the heat function. The heat function is defined through the relation
\beq
\mathfrak{H} = E + \sum_{\{\ell\}} x_{\ell} X_{\ell},
\label{hf_def}
\eeq
where $E$ is the internal energy and the $x$ is an intensive thermodynamic variable whose conjugate extensive 
variable is $X$. Defining $\mathfrak{X}_{1}$ and $\mathfrak{X}_{2}$ as the variables corresponding to the chemical 
and mechanical equilibrium the specific form of each ensemble, its heat function and the corresponding entropy are
listed in the Table 1.
\begin{table}[!ht]
{\small
\begin{center}
\caption{List of Adiabatic ensembles}
\vspace{0.2cm}
\begin{tabular}{|c|c|c|c|}
\hline
   &         & Heat function & Entropy \\
$\ell$ &Ensemble & $\mathfrak{H} = E + \displaystyle{\sum_{\{\ell\}}}\, x_{\ell} X_{\ell}$  &
$S(\mathfrak{X}_{1},\mathfrak{X}_{2},\mathfrak{H})$\\
\hline 
$0$ & Microcanonical & $E$ & $S(N,V,E)$\\
& $(N,V,E)$      & (Internal energy) &  \\
\hline
$1$ & Isoenthalpic - isobaric &  $H = E + P V$ & $S(N,P,H)$\\
& $(N,P,H)$      & (Enthalpy) & \\
\hline
$1$ & Third adiabatic ensemble &  $\mathsf{L} = E - \mu N $ & $S(\mu,V,{\mathsf{L}})$\\
 & $(\mu,V,\mathsf{L})$      & (Hill energy) & \\
\hline
$2$ & Fourth adiabatic ensemble &  $\mathsf{R} = E + P V - \mu N$ & $S(\mu,P,{\mathsf{R}})$\\
& $(\mu,P,\mathsf{R})$      & (Ray energy) & \\
\hline
\end{tabular}
\label{adiabatic_tab}
\end{center}
}
\end{table}

Of the four adiabatic ensembles, the microcanonical and the isoenthalpic-isobaric ensembles
described in [\cite{EG}-\cite{JR1}] are well known, but the other two ensembles namely $(\mu,V,\mathsf{L})$ and $(\mu,P,\mathsf{R})$ 
introduced through References [\cite{JR2},\cite{JR4}] are relatively less known.  But they are important 
in studying adiabatically confined systems with variable number of particles.  
For example in Ref. [\cite{JR93}] a Monte Carlo simulation 
of a system of liquid palladium has been carried out in the $(\mu,P,\mathsf{R})$ ensemble.
The simulation was much more convenient in this ensemble and the results agreed with those
obtained in the grandcanonical ensemble. Realizing the importance of classifying the ensembles,
a unified treatment of the adiabatic ensembles was carried out in [\cite{HG1}].  Later on the
physical realizations corresponding to the eight ensembles and their interrelations through
Laplace-Legendre transforms was studied in [\cite{HG2}].

Tsallis proposed a generalization of the existing Boltzmann Gibbs statistical mechanics [\cite{CT1}] through 
the introduction of $q$-deformed logarithm and exponential functions. The entropic expression in the generalized
statistical mechanics was based on the $q$-deformed logarithm. Many other deformed entropies like the $\kappa$-entropy
[\cite{GK2001}], the basic deformed entropies [\cite{AL1}] etc., were also proposed.  The equilibrium formulation
of statistical mechanics based on the deformed entropies were developed and applied to a wide variety of systems
like long range interacting rotators [\cite{BJC02}], relativistic gases [\cite{AL2},\cite{CCN2}], and systems with long range microscopic
memory [\cite{MT12}].  Classification of the eight ensembles in to two different classes namely, the isothermal
and the adiabatic class, and a unified description for each class in the framework of generalized statistical mechanics
based on Tsallis $q$ entropy has been carried out in [\cite{CM2011}]. 

Later it was shown in [\cite{SW05}] that an information theoretic entropy known as the 
Sharma-Taneja-Mittal entropy [\cite{ST75},\cite{M75}] based on a two parameter logarithm was a natural 
generalization of both the Tsallis $q$-entropy and the $\kappa$-entropy. A study of nonlinear Fokker Planck
equation corresponding to the two parameter $(r,s)$ Sharma-Taneja-Mittal entropy was carried out in [\cite{TF99}] and the 
role played by the two parameters was investigated. One of the parameter `$r$' was found to describe
the nature of the stationary solution i.e., whether it is a weak stationary or a strong stationary solution and
hence $r$ determines the degree of distortion of the usually obtained Gaussian distribution.  The other parameter
`$s$' characterizes the dynamical properties of the transient solution and thus distinctly differentiates the 
regions of subdiffusion and superdiffusion. Investigation using the Lie symmetries [\cite{AS09}] proved that the localized
initial states are well approximated by the two parameter Gaussian rather than the standard Gaussian. The Sharma-Taneja-Mittal
entropy is nonsymmetric with respect to its parameters, so with a view to 
provide a symmetric generalization of the Tsallis $q$ entropy, a two parameter logarithm and its inverse function the two 
parameter exponential were introduced in [\cite{S2007}]. The doubly deformed $(q,q^{\prime})$ logarithm and its 
inverse the $(q,q^{\prime})$-exponential for a real variable $x$ are
\bea
\ln_{q,q^{\prime}} (x) &=& \frac{1}{1-q^{\prime}} \left[\exp \left(\frac{1-q^{\prime}}{1-q} (x^{1-q} -1)\right) -1\right], 
\label{2plog}\\
\exp_{q,q^{\prime}} (x) &=& \left[ 1 + \frac{1-q}{1-q^{\prime}} \ln\left(1+(1-q^{\prime})x \right) \right]^\frac{1}{1-q}.
\label{2pexp}
\eea
The two parameter algebra based on the doubly deformed logarithm and the exponential was found [\cite{EB08}] to be 
nondistributive in nature. Based on the definition of the two parameter logarithm $(\ref{2plog})$ the generalized entropy is
\bea
S_{q,q^{\prime}} &=& k_{{}_{TS}} \, \sum_{i=1}^{w} p_{i} \, \ln_{q,q^{\prime}} \left(\frac{1}{p_{i}}\right), \nn \\
&=& \frac{k_{{}_{TS}}}{1-q^{\prime}} \, \sum_{i=1}^{w} p_{i} \, 
    \left(\exp \left(\frac{1-q^{\prime}}{1-q} (p_{i}^{q-1} -1)\right) -1 \right).
\label{2p_entropy}
\eea
Under the assumption of equiprobability i.e., $p_{i} = 1/w$ the two parameter entropy (\ref{2p_entropy}) reads:
\beq
S_{q,q^{\prime}} = \frac{k_{{}_{TS}}}{1-q^{\prime}} \, \left(\exp \left(\frac{1-q^{\prime}}{1-q} (w^{1-q} -1)\right) -1 \right).
\label{2p_entropy_mc}
\eeq
Both the doubly deformed logarithm (\ref{2plog}) and the exponential (\ref{2pexp}) 
reduce to the respective one parameter logarithm or exponential defined in [\cite{CT1}], when we let either 
$q$ or $q^{\prime}$ tend to unity.  Due to this property, the Tsallis generalized entropy can be recovered by allowing
either of the deformation parameters to approach it limiting value.   

Nonextensivity has been proved to introduce effective interactions in ideal systems. In [\cite{CCN1}],
we observed that an effective interaction was introduced between different degrees of freedom 
in an ideal diatomic gas molecule and in Ref. [\cite{AL1}] it was clearly proved that the effect of basic 
deformation of a free ideal gas can be viewed as an effective interaction described by an 
Hamiltonian with logarithmic interactions. Recent simulational studies [\cite{RGD09}] show that classical 
collisions of a single ion in a radio frequency ion trap with an ideal gas generates non-Maxwellian 
distribution functions.  Thus nonextensive statistical mechanics may be considered 
as a method to study interacting systems using noninteracting Hamiltonians, wherein the interactions 
are introduced through the nonextensivity in statistical mechanics. Taking this point of view, we feel 
that using a two parameter entropy gives us more freedom in introducing different forms of 
interactions both in terms of variety and their strengths. Throughout this article we take this spirit 
in investigating the ideal systems.  Investigating the effect of a gravitational field on a system of 
ideal gas and one dimensional hard rods will help us in understanding the motion of a interacting gas cloud under the 
influence of gravity.  We expect these calculations to correspond to some scenarios in which a massive object
gravitationally attracts gas molecules from a gas cloud.

A canonical formulation of generalized statistical mechanics based on the two parameter entropy (\ref{2p_entropy}) was
carried out in Ref. [\cite{As2008}] and the authors found that the distribution could be obtained in terms
of the Lambert's $W$-function.  This can be formally extended to describe the other members of the isothermal class of ensembles.
In our current work we investigate the adiabatic class of ensembles of the generalized statistical mechanics
based on the two parameter entropy.  We provide a unified description of the adiabatic class and provide a
two parameter generalization of the equipartition theorem, virial theorem and the adiabatic theorem.  All the 
four ensembles are illustrated using the classical nonrelativistic ideal gas and their corresponding heat functions
and heat capacities are obtained analytically. For the ideal gas, the phase space volume in the microcanonical and the
isoenthalpic-isobaric ensemble can be found easily. To compute the respective heat function we need to adopt the 
large $N$ limit and the results are obtained in terms of the Lambert's $W$-function. A brief introduction to the
$W$-function is provided in Section II of this article. The phase space volume corresponding to the $(\mu,V,\mathsf{L})$
and the $(\mu,P,\mathsf{R})$ ensembles could not be evaluated exactly and we overcome this
by using the large $N$ limit. Since the large $N$ limit has been already used in the computation of the phase space volume
no further approximations are needed in the calculation of the heat function and the specific heat. The effect of
gravity on a classical nonrelativistic ideal gas, and a system of one dimensional hard rod gas has been investigated 
[\cite{Ro1997},\cite{Ro1999}] in the extensive Boltzmann-Gibbs statistical mechanics. The authors found that the entropy is
a decreasing function of the gravitational field indicating that gravity has an ordering effect in a thermodynamic system. 
In the present work we study the effect of a gravitational field in the framework of two parameter $(q,q^{\prime})$-entropy. 
First a $D$-dimensional nonrelativistic classical ideal gas confined in a finite region of space, subjected to 
an external gravitational field is investigated.  The entropy and the specific heat are obtained exactly as a function of the internal 
energy.  In the infinite height limit, the internal energy could be found only when we assume that the number of 
particles is very large. Next we study a system of one dimensional hard rods in the presence of gravity.  Analogous to 
the classical ideal gas, the entropy and the specific heat are obtained as a function of internal energy.  But in the 
infinite height limit the internal energy could be obtained as a function of temperature subject to the condition that
the number of particles is very large.  The heat capacities were found to admit both positive and negative values, in contrast 
to the previous results in [\cite{SA99}] where the specific heat of a system of gas molecules in thermodynamic limit permits only 
negative values.  Specific heats are supposed to characterize the amount of energy change when temperature is varied.  Energy can both
be liberated or absorbed depending on the interactions in the system as $T$ is varied.  Our contention here is that with 
the $(q,q^{\prime})$ dependence the system exhibits effective interactions.  The fact that specific heat can become negative 
for some range of $(q,q^{\prime})$ and may give clues about {\it(i)} Choosing the range of $(q,q^{\prime})$ for systems described
by positive specific heat.  {\it(ii)} If for a non-Boltzmann system with negative specific heat, it can help us to infer
about interacting systems.  

The plan of the article is as follows: Following the introduction in Section I we give a brief summary of the 
Lambert's $W$ function in Section II.  A unified description of the thermostatistical structure of all the four adiabatic
ensembles is provided in Section III.  In Section IV, we study all the four ensembles using classical nonrelativistic 
ideal gas, and each ensemble is treated separately in a subsection. The effect of an external 
gravitational field on thermodynamic system is studied in Section V.  The first part of the section deals with the 
effect of gravity on a $D$-dimensional classical ideal gas.  A system of one dimensional hard rod gas is examined in the
second part of the section.  The investigations are carried out in the microcanonical ensemble. We present our conclusions
in Section V. 
%
%
%
%
%
\setcounter{equation}{0}
\section{Lambert's W-function: a primer} 
\label{wfunction}
A brief summary of the Lambert's $W$-function and its mathematical properties is presented
in the current section.  Lambert's $W$-function represented by $W(z)$ is defined as the multivalued 
inverse of the function $w \ e^{w} = z$ and satisfies the relation
\beq
W(z) \ e^{W(z)} = z,  \qquad z \in \mathbb{C}. 
\label{Wfunc_def}   
\eeq 
The various branches corresponding to the $W$-function are indexed by $k=0,\pm1,\pm2,...,$
and for real $z<-1/e$, the function $W(z)$ is always complex and multivalued. For real 
$z$ in the range $-1/e \leq z < 0$, the function $W(z)$ comprises of two real branches namely
$W_{0}(z)$ and $W_{-1}(z)$ represented by the solid line and the dotted line respectively in 
Figure 1.  The branch $W_{0}(z)$ satisfies the condition $W(z) \geq -1$,
and is generally known as the principal branch of the $W$ function. Correspondingly, when 
$W(z) \leq -1$ we have the $W_{-1}(z)$ branch.  
\begin{figure}
\begin{center}
 \scalebox{1.0}{\input{wfunction.tex}}
\label{wfunc} \\
Figure 1: A plot of the Lambert's $W$ function $W(x)$ as a function of $x$. 
\end{center}
\end{figure}
The principal branch is analytic at $z=0$ and has the series expansion
\beq
W_{0}(z) = \sum_{n=1}^{\infty} \frac{(-n)^{n-1}}{n!} \, z^{n},
\eeq
with a radius of convergence $e^{-1}$.  Similarly the series corresponding
to the branch $W_{-1}$  is
\beq
W_{-1}(z) = \sum_{n=0}^{\infty} c_{n} \, f^{n}(z); 
\qquad \qquad f(z) = \sqrt{2 \ (ez + 1)},
\eeq
where the expansion coefficients can be computed from the recurrence relations
\bea
c_{k} &=& \frac{k-1}{k+1} \,\left(\frac{c_{k-2}}{2} + \frac{d_{k-2}}{4} \right)
          - \frac{d_{k}}{2} - \frac{c_{k-1}}{k+1},
\qquad 
d_{k} = \sum_{j=2}^{k-1} c_{j} \, c_{k+1-j},
\label{exp_coeff}  \\
c_{0} &=& -1, \qquad c_{1} = 1, \qquad  d_{0} = 2, \qquad d_{1} = -1.
\label{ini_coeff}
\eea
The series converges for $-1/e \leq z < 0$, which covers the whole domain
of the $W_{-1}$ branch. From (\ref{Wfunc_def}) the derivative of the $W$
function is found
\beq
W^{\prime}(z) = \frac{W(z)}{z \ (1+W(z))}. 
\eeq
In the current article physical requirements restricts our choice of the $W$ function to the principal branch. 
An excellent introduction to the Lambert's $W$-function and its applications to engineering
problems is discussed in [\cite{C96}]. Recently in signal processing the role of the $W_{-1}$ 
branch was analyzed [\cite{FCB02}].  The Lambert's function has been used in the study of 
both classical [\cite{Ca2004}] and quantum statistical mechanics [\cite{Va2009},\cite{Va2010}].  
It also occurs [\cite{EL05}] in the study of Fokker Planck equation in the small noise limit. 
%
%
%
%
\setcounter{equation}{0}
\section{Adiabatic ensemble: Generalized formulation} 
\label{micro}
The individual members of an adiabatic ensemble have the same value of the heat
function though they can be at different temperatures. For a system in 
thermodynamic equilibrium, there are four different adiabatic ensembles namely
the microcanonical ensemble $(N,V,U)$, the isoenthalpic-isobaric ensemble 
$(N,P,H)$, the adiabatic ensemble with number fluctuations $(\mu,V,\mathsf{L})$, 
and the adiabatic ensemble with both number and volume fluctuations $(\mu,P,\mathsf{R})$.  
The microstate of a system of $N$ particles can be represented by a single point in
the $2 DN$ dimensional phase space. Corresponding to a particular value of the heat 
function which is a macrostate, we have a huge number of microstates. We need to compute
the total number of microstates since it is a measure of the entropy.  The points 
denoting the microstate of the system lie so close to each  other that the surface area
of the constant heat function curve in the phase space is regarded as a measure of the 
total number of microstates. For a system described by a Hamiltonian $\mathcal{H}$ the
surface area corresponding to a constant heat function
$\mathfrak{H}$ curve can be calculated from 
\bea
\Omega (\mathfrak{X}_{1},\mathfrak{X}_{2},\mathfrak{H})
              = \sum_{X_{\{\ell\}}} \, \frac{1}{N! \, h^{DN}}
                \int_{r_{i}} \int_{p_{i}}
                \delta \Big(\mathcal{H}+\sum_{\{\ell\}}
                x_{\ell}X_{\ell} - {\mathfrak{H}}\Big)
                \prod_{i=1}^{N} {\rm d}^{D} r_{i} \, {\rm d}^{D} p_{i}.
\label{sds_en_TS}
\eea
Similarly the phase space volume enclosed by the constant heat function ${\mathfrak{H}}$ curve is
\beq
\varSigma(\mathfrak{X}_{1},\mathfrak{X}_{2},\mathfrak{H}) =
                   \sum_{X_{\{\ell\}}} \, \frac{1}{ N! \, h^{DN}}
                   \int_{r_{i} } \int_{p_{i}}
                   \Theta \Big(\mathcal{H} + \sum_{\{\ell\}}
                   x_{\ell}X_{\ell} - {\mathfrak{H}} \Big)
                   \prod_{i=1}^{N} {\rm d}^{D} r_{i}\,  {\rm d}^{D} p_{i},
\label{vds_en_TS}
\eeq
where $\Theta$ is the Heaviside step function.
Computation of the volume of the constant heat function curve assumes significance because 
of the difficulty in calculating the area of the curve. The volume enclosed by the phase space curve 
and its surface area are related via the expression
\beq
\Omega (\mathfrak{X}_{1},\mathfrak{X}_{2},\mathfrak{H}) = \frac{\partial}{\partial \mathfrak{H}} 
                                                          \varSigma(\mathfrak{X}_{1},\mathfrak{X}_{2},\mathfrak{H}).
\label{sds_vds_rel}
\eeq
Since the phase space volume  is a measure of the number of the microstates of the system, 
the two parameter entropy can be directly obtained from the knowledge of 
$\varSigma(\mathfrak{X}_{1},\mathfrak{X}_{2},\mathfrak{H})$ via the relation
\beq
S_{q,q^{\prime}}(\mathfrak{X}_{1},\mathfrak{X}_{2},\mathfrak{H}) = k \; \ln_{q,q^{\prime}} 
                                                                  \varSigma(\mathfrak{X}_{1},\mathfrak{X}_{2},\mathfrak{H}).
\label{entr_def_rel}
\eeq
For a given adiabatic ensemble the temperature is defined via the relation
\beq
T = \left(\frac{\partial S_{q,q^{\prime}}}{\partial {\mathfrak{H}}}\right)^{-1}
  = \frac{(\varSigma(\mathfrak{X}_{1},\mathfrak{X}_{2},\mathfrak{H}))^{{}^{q}}}
    {k  \;\Omega(\mathfrak{X}_{1},\mathfrak{X}_{2},\mathfrak{H}) \, 
    \exp\left(\frac{1-q^{\prime}}{1-q}((\varSigma(\mathfrak{X}_{1},\mathfrak{X}_{2},\mathfrak{H}))^{1-q} -1) \right)}.
\label{temp_def_TS}
\eeq
Using the definition of the temperature (\ref{temp_def_TS}), the phase space volume (\ref{vds_en_TS}) and 
the surface area (\ref{sds_en_TS}), we can calculate the expression corresponding to 
the heat function.  From the expression for the heat function the specific heat can be calculated through 
the relation
\beq
C_{x_{\{\ell\}}} = \frac{\partial {\mathfrak{H}}_{q}}
                 {\partial T} \bigg|_{x_{\{\ell\}}}.
\label{psh_def_TS}
\eeq
For any adiabatic ensemble the expectation value of an observable $O$ is defined as
\beq
\langle O \rangle = \frac{1}{\Omega(\mathfrak{X}_{1},\mathfrak{X}_{2},\mathfrak{H})} \,
                    \sum_{X_{\{\ell\}}}
                    \frac{1}{N! \, h^{DN}} \,
                    \int_{r_{i} } \int_{p_{i}} O \;\;
                    \delta \Big(\mathcal{H} + \sum_{\{\ell\}}
                    x_{\ell} X_{\ell} - {\mathfrak{H}}\Big)
                    \prod_{i=1}^{N} {\rm d}^{D} r_{i} \, {\rm d}^{D} p_{i}.
\label{oex_sds_TS}
\eeq
The above expression is extremely useful in computing the average energy in the $(N,P,H)$, 
$(\mu,V,\mathsf{L})$ and the $(\mu,P,\mathsf{R})$ ensembles.  By using the suitable Legendre
transformations, the average energy can also be obtained from the heat functions corresponding to 
these ensembles. 
From the entropy the extensive thermodynamic quantities whose conjugate intensive variables are held fixed 
can be computed through the relation
\beq
X_{\ell}(\mathfrak{X}_{1},\mathfrak{X}_{2},\mathfrak{H}) = - \frac{1}{\beta} \;
                                                               \frac{\partial}{\partial x_{\ell}}\,
                                                               S(\mathfrak{X}_{1},\mathfrak{X}_{2},\mathfrak{H}).
\label{vol_N_def_TS}
\eeq
The equipartition theorem is derived in a unified manner for all the four adiabatic ensembles in the framework
of generalized statistical mechanics based on the Schw\"{a}mmle-Tsallis $(q,q^{\prime})$ entropy. Representing the 
phase space variables $r_{i}$ and $p_{i}$ $(i=1,2,...,3N)$ by a common variable $y_{i}$, we find the 
following expectation value 
\beq
\bigg\langle y_{i}\; \frac{\partial {\mathcal{H}}}
{\partial y_{j}} \bigg\rangle = k T \;
                                 \left(\varSigma(\mathfrak{X}_{1},\mathfrak{X}_{2},\mathfrak{H})\right)^{1-q} \; 
                                 \exp\left(\frac{1-q^{\prime}}{1-q}((\varSigma(\mathfrak{X}_{1},\mathfrak{X}_{2},\mathfrak{H}))^{1-q} -1) \right)
                                 \delta_{ij},
\label{ept_uf_TS}
\eeq
which is a generalization of the equipartition function. We notice that the $(q,q^{\prime})$ generalized 
equipartition theorem is dependent on the phase space volume in contrast to the extensive Boltzmann-Gibbs statistics. 
This observation is in line with the earlier result obtained for the Tsallis entropy in Ref. [\cite{CM2011}] and 
yields the Boltzmann-Gibbs limit when both the nonextensive parameters $q$ and  $q^{\prime}$ are set to unity. 
When we let the variable $y_{i}$ to be the coordinate $r_{i}$ and invoke the Dirac delta function, we get a 
specific form of the equipartition theorem:
\bea
\bigg\langle r_{i}\; \frac{\partial {\mathcal{H}}}
{\partial r_{i}} \bigg\rangle \equiv - \langle r_{i} \, \dot{p}_{i} \rangle 
                                 = \langle r_{i} \, F_{i} \rangle
                                &=&  k T \;  \left(\varSigma(\mathfrak{X}_{1},\mathfrak{X}_{2},\mathfrak{H})\right)^{1-q} \nn \\
                                & & \exp\left(\frac{1-q^{\prime}}{1-q}((\varSigma(\mathfrak{X}_{1},\mathfrak{X}_{2},\mathfrak{H}))^{1-q} -1) \right).
\label{ept_pos_TS}
\eea
Similarly when we set $y_{i}$ to be equal to $p_{i}$ the momentum variable we obtain
\beq
\bigg\langle p_{i}\, \frac{\partial {\mathcal{H}}}{\partial p_{i}} \bigg\rangle \equiv
\langle p_{i} \, \dot{r}_{i} \rangle = k T \, \left(\varSigma(\mathfrak{X}_{1},\mathfrak{X}_{2},\mathfrak{H})\right)^{1-q} \, 
                                      \exp\left(\frac{1-q^{\prime}}{1-q}((\varSigma(\mathfrak{X}_{1},\mathfrak{X}_{2},\mathfrak{H}))^{1-q} -1) \right).
\label{ept_mtm_TS}
\eeq
Through a canonical transformation, sometimes the Hamiltonian of a system can be written in the following form
$\mathcal{H} = \sum_{i} (A_{i} \ P_{i}^{2} + B_{i} \ Q_{i}^{2})$. Aided by equations (\ref{ept_pos_TS}) and 
(\ref{ept_mtm_TS}) we find the expectation value of such an Hamiltonian to be
\beq
\langle \mathcal{H} \rangle = DN \, k T\, \left(\varSigma(\mathfrak{X}_{1},\mathfrak{X}_{2},\mathfrak{H})\right)^{1-q} \,
                              \exp\left(\frac{1-q^{\prime}}{1-q}((\varSigma(\mathfrak{X}_{1},\mathfrak{X}_{2},
                              \mathfrak{H}))^{1-q} -1) \right).                        
\eeq
A two parametric generalization of the virial theorem could be obtained from the relation (\ref{ept_pos_TS}) 
and reads: 
\beq
\bigg \langle \sum_{i}^{DN} r_{i} \, \dot{p}_{i} \bigg \rangle = - DN \, k T \,
                                                                  \left(\varSigma(\mathfrak{X}_{1},\mathfrak{X}_{2},
									\mathfrak{H})\right)^{1-q} \, 
                                                                  \exp\left(\frac{1-q^{\prime}}{1-q}((\varSigma(\mathfrak{X}_{1},
                                                                  \mathfrak{X}_{2},\mathfrak{H}))^{1-q} -1) \right).
\label{vt_TS}
\eeq
Below we verify the adiabatic theorem for the generalized statistical mechanics based on the 
two parameter entropy.  Let us consider the Hamiltonian to contain an external parameter
$a$, in addition to the phase space co-ordinates.  An expectation value of the derivative of 
the Hamiltonian with respect to the external parameter yields the thermodynamic conjugate 
variable $f$ corresponding to the parameter $a$
\bea
\bigg\langle \frac{\partial \mathcal{H}} {\partial a} \bigg\rangle &=&
		\frac{1}{\Omega(\mathfrak{X}_{1},\mathfrak{X}_{2},\mathfrak{H})}
		\sum_{X_{\ell}} \frac{1}{N! \, h^{DN}} \, \int_{r_{i}} \int_{p_{i}}
		\left(\frac{\partial \mathcal{H}}{\partial a} \right)
		\delta(\mathcal{H} + \sum_{X_{\ell}} x_{\ell} X_{\ell} - \mathfrak{H})
		\prod_{i=1}^{N} d^{D}r_{i} \, d^{D} p_{i}, \nn \\
		&=&\frac{\partial \mathfrak{H}} {\partial a} = f.
\label{adb_thr_TS}
\eea
The description of the adiabatic ensembles given above has been illustrated through 
certain examples. First we consider a classical ideal gas and provide an analytic 
solution to the specific heat in all the four ensembles by using a large $N$ 
approximation.  Later on we study a microcanonical $D$-dimensional classical 
ideal gas under the influence of gravity and compute its specific heat.  Finally
we investigate a hard rod gas under gravity in the framework of microcanonical ensemble. 
%
%
%
%
\setcounter{equation}{0}
\section{Application - Classical ideal gas}
\label{Ideal gas}
The Hamiltonian of a nonrelativistic classical ideal gas in $D$ dimensions is
\beq
\mathcal{H} = \sum_{i} \frac{p_{i}^{2}}{2m},    \qquad
p_{i}=|{\bf{p}}_{i}|, 
\label{ham_ig}
\eeq
where ${\bf{p}}_{i}$ for $(i = 1,2,...,N)$ represent the $D$-dimensional momenta of the gas molecules.
In this section we find the phase space volume corresponding to this Hamiltonian for all the 
four ensembles.  From the phase space volume we derive 
the relevant thermodynamic quantities like the entropy, the heat function and 
the heat capacity.  
\subsection{Microcanonical ensemble}
The classical nonrelativistic ideal gas described by (\ref{ham_ig}) is studied in the microcanonical
ensemble.  In order to compute the entropy of the system, we calculate the phase space volume 
enclosed by the constant energy curve.  Substituting the expression of the Hamiltonian in 
(\ref{vds_en_TS})
\beq
\varSigma(N,V,E) =  \frac{1}{N! \, h^{DN}} \; \int_{r_{i}} \int_{p_{i}}
                   \Theta\left(\sum_{i} \frac{p_{i}^{2}}{2m} - E\right)\;
                   \prod_{i=1}^{N} {\rm d}^{D} r_{i}\, {\rm d}^{D} p_{i}.
\label{mc_vs}
\eeq
The phase space integral is computed in the following manner:  First we notice that 
the momentum integration is the volume of a $DN$-dimensional sphere of radius
$\sqrt{2mE}$.  Next we integrate over the position co-ordinates to obtain the 
phase space volume in the microcanonical ensemble
\beq
\varSigma(N,V,E) = \frac{V^{N}}{N!} \; \frac{\mathcal{M}^{N}}{\Gamma\left(\frac{D N}{2}+1\right)} \;
                    E^{\frac{DN}{2}},
\label{mc_vs_fe}
\eeq
where we define $\mathcal{M} = (2 \pi m/h^{2})^{D/2}$ for the sake of convenience. 
The corresponding surface area enclosed by the phase space curve is 
\beq
\Omega(N,V,E) = \frac{V^{N}}{N!} \; \frac{\mathcal{M}^{N}}{\Gamma\left(\frac{D N}{2}\right)} \;
                E^{\frac{DN}{2}-1}.
\eeq
The microcanonical entropy of a classical ideal gas obtained from the knowledge of the 
phase space volume (\ref{mc_vs_fe}) reads: 
\beq
S_{q,q^{\prime}} = \frac{k}{1-q^{\prime}} \; \left[\exp \left(\frac{1-q^{\prime}}{1-q} 
                   \big((\Xi_{mc} \; E^{\frac{DN}{2}})^{1-q} -1\big)\right) -1\right],
\label{entr_mc_ig} 
\eeq
where the factor $\Xi_{mc}$ is defined as
\beq
\Xi_{mc} = \frac{V^{N}}{N!} \; \frac{\mathcal{M}^{N}}{\Gamma\left(\frac{D N}{2}+1\right)}.
\label{xi_def_ig}
\eeq
In the limit $(q,q^{\prime}) \rightarrow 1$, we recover the extensive Boltzmann Gibbs entropy. 
From the definition of temperature (\ref{temp_def_TS}), we arrive at the following expression
\beq
\frac{1}{T} = \frac{DN}{2} \, k \; \Xi_{mc}^{1-q} \, E^{(1-q)\,\frac{DN}{2}  - 1} \,
                   \exp \left(\frac{1-q^{\prime}}{1-q} 
                   \big((\Xi_{mc} \; E^{\frac{DN}{2}})^{1-q} -1\big)\right).
\label{temp_inte_rel}
\eeq
Inversion of the above relation (\ref{temp_inte_rel}), to obtain the internal energy is intractable and so,
we look into the naturally occurring large $N$ limit.  In this limit we can safely neglect the factor of one 
in comparison with $(1-q) DN/2$ and this enables us to make the following approximation 
$E^{(1-q) \, {\frac{DN}{2}} -1} \approx E^{(1-q) \, {\frac{DN}{2}}}$. For very large values of $N$ the 
deformation parameter $(1-q)$ should be $O(1/N)$ for the factor $1$ to make a reasonable contribution. 
Using the assumption outlined above we get 
\beq
\exp\left( \frac{1-q^{\prime}}{1-q} \right) \frac{2}{DN \, k \, T} = 
\Xi_{mc}^{1-q} \, E^{(1-q) \,\frac{DN}{2}} \, 
\exp\left(\frac{1-q^{\prime}}{1-q} \, (\Xi_{mc} \; E^{\frac{DN}{2}})^{1-q} \right). 
\label{lamb_eqn}
\eeq
The inversion of the above function leads to an expression of the internal energy in terms of the 
Lambert's $W$ function as
\beq
E = \left[\frac{1-q}{1-q^{\prime}} \; \frac{1}{\Xi_{mc}^{1-q}} \; 
    W_{0} \left(\frac{1-q^{\prime}}{1-q} \, \exp\left( \frac{1-q^{\prime}}{1-q}\right) 
    \frac{2 \beta}{DN} \right)\right]^
    {\frac{2}{(1-q)\, DN}}.
\label{mc_ig_fe}
\eeq
The requirement that the entropy be concave decides the range of the deformation parameters. From
the discussion in Ref. [\cite{S2007}], we notice that the entropy is concave in the region $q + q^{\prime} \geq 1$
excluding the region $0 \leq (q,q^{\prime}) \leq 1$, where the entropy does not have fixed curvature sign. The argument
of the $W$ function in (\ref{mc_ig_fe}) is positive in the region (a) for which $q > 1$ and $q^{\prime} > 1$, and hence the energy
in this region is characterized by the principal branch. Under the restriction $q + q^{\prime} < 1$, there are 
two other major regions namely, {(b)} $q > 1$ and $-\infty < q^{\prime} < 1$  and {(c)} $q^{\prime} > 1$ 
and $-\infty < q < 1$ in which the argument of the $W$ function is negative. Though we have a choice between the
$W_{0}$ and the $W_{-1}$ branch, the continuity requirement on energy restricts our choice to the $W_{0}$ branch.
Thus we conclude that only the principal branch contributes in the definition of energy. In all the subsequent 
discussions in this article, we maintain the definition of the regions (a), (b) and (c) as described above.  
The specific heat at constant volume evaluated from the internal energy is 
\beq
C_{V} \equiv \frac{\partial E}{\partial T} = - \frac{k \, \beta}{(1-q) \, DN/2} \; 
            \frac{W_{0} (\mathsf{b} \, \beta)}{1 + W_{0} (\mathsf{b} \, \beta)} \;
          \left[\mathsf{a}\,  W_{0} (\mathsf{b} \, \beta)\right]^{\frac{1-(1-q)DN/2}{(1-q) DN/2}}, 
\label{mc_ig_spht}
\eeq
where the factors $\mathsf{a}$ and $\mathsf{b}$ are as defined below:
\beq
\mathsf{a} = \frac{1-q}{1-q^{\prime}} \; \frac{1}{\Xi_{mc}^{1-q}}, \qquad
\mathsf{b} = \frac{1-q^{\prime}}{1-q} \, \exp\left(\frac{1-q^{\prime}}{1-q}\right) 
             \frac{2}{DN}.
\label{mc_ab_def}
\eeq
Invoking the large $N$ limit in the final expression of the specific heat we arrive at 
\beq
C_{V} = - \frac{k \, \beta}{(1-q) \, DN/2} \; \frac{1}{\mathsf{a}(1 + W_{0} (\mathsf{b} \, \beta))}.
\label{mc_ig_spht_app}
\eeq
From the expression of the specific heat (\ref{mc_ig_spht_app}), we notice that it can be either 
positive or negative depending on the values of the deformation parameters $q$ and $q^{\prime}$. In the
region (a) and region (c) the specific heat is positive, whereas in the region (b) it is negative. 
We do not recover the Boltzmann Gibbs statistics from our above calculations, because in the 
computation of energy (\ref{mc_ig_fe}), we have made use of the large $N$ limit which does not commute with extensive
$(q,q^{\prime}) \rightarrow 1$ limit. 
\subsection{Isoenthalpic-isobaric ensemble}
A system which exchanges internal energy and volume with its surroundings in such a way that its
enthalpy remains constant is described by the isoenthalpic-isobaric ensemble. In order to calculate
the thermodynamic quantities we first find the phase space volume enclosed by the constant 
enthalpy curve. The integral expression corresponding to the phase space volume is 
\beq
\varSigma(N,P,H) = \sum_{V} \frac{1}{N! \, h^{DN}} \; \int_{r_{i}} \int_{p_{i}}
                   \Theta\left(\sum_{i} \frac{p_{i}^{2}}{2m} + P V - H\right)\;
                   \prod_{i=1}^{N} {\rm d}^{D} r_{i}\, {\rm d}^{D} p_{i}.
\label{VS_nrg}
\eeq
Using the fact that the momentum integration is nothing but geometrically equivalent to a 
$DN$ dimensional sphere of radius $\sqrt{2m(H - PV)}$, and integrating over the 
position co-ordinates we arrive at 
\beq
\varSigma(N,P,H) = \frac{\mathcal{M}^{N}}{N!} \; \frac{1}{\Gamma\left(\frac{D N}{2}+1\right)} \;
                   \sum_{V} V^{N} (H-P V)^{\frac{DN}{2}}.
\label{V_sum}
\eeq
In the next step we consider a summation over the volume eigenstates.  Since the volume states
are very closely spaced the summation is replaced by an integration.  But an integral over the
volume leads to over counting of the eigenstates.  To overcome this we employ the shell particle 
method of counting volume states developed in [\cite{DC1},\cite{DC2}].  In this method we take 
into account only the minimum volume needed to confine a particular configuration.  The minimum 
volume needed to confine a particular configuration is found by imposing a condition wherein we 
require atleast one particle to lie on the boundary of the system.  All the equivalent ways of 
choosing a minimum volume for a particular configuration is treated as the same volume eigenstate 
and is considered only once.  Using this shell particle technique to reject the redundant volume 
states we arrive at 
\beq
\varSigma(N,P,H) = \mathcal{M}^{N} \;
               \left(\frac{1}{P}\right)^{N} \;
               \frac{H^{\mathfrak{D}}}
               { \Gamma({\mathfrak{D}} + 1 )}, \qquad
\mathfrak{D} = \frac{DN}{2} + N.               
\label{vds_nr_fe}
\eeq
A similar evaluation of the surface area enclosed by the curve of constant enthalpy yields
\beq
\Omega(N,P,H) = \mathcal{M}^{N} \;
            \left(\frac{1}{P}\right)^{N} \;
            \frac{H^{{\mathfrak{D}}-1}}
            {\Gamma({\mathfrak{D}})}.
\label{sds_nr_fe}
\eeq
From the knowledge of the phase space volume (\ref{vds_nr_fe}), the entropy of a classical ideal gas in the 
isoenthalpic-isobaric ensemble is found: 
\beq
S_{q,q^{\prime}} = \frac{k}{1-q^{\prime}} \; \left[\exp \left(\frac{1-q^{\prime}}{1-q} 
                   \big((\Xi_{ie} \; {H^{\mathfrak{D}}})^{1-q} -1\big)\right) -1\right].
\label{entr_ie_ig} 
\eeq
The Boltzmann Gibbs entropy is recovered in the limit $(q,q^{\prime}) \rightarrow 1$. 
The factor $\Xi_{ie}$ used in the above equation is 
\beq
\Xi_{ie} = \mathcal{M}^{N} \;
               \left(\frac{1}{P}\right)^{N} \;
               \frac{1}{ \Gamma({\mathfrak{D}} + 1)}. 
\eeq
The partial derivative of the entropy (\ref{entr_ie_ig}) with respect to the enthalpy gives the temperature of the 
isoenthalpic-isobaric ensemble
\beq
\frac{1}{T} = {\mathfrak{D}} \, k \, \Xi_{ie}^{1-q} \, H^{(1-q)\, {\mathfrak{D}} - 1} \,
                   \exp \left(\frac{1-q^{\prime}}{1-q} 
                   \big((\Xi_{ie} \; {H^{\mathfrak{D}}})^{1-q} -1\big)\right).
\label{temp_def_ie}
\eeq
In order to invert the above equation we assume the large $N$ limit and this consequently 
leads to the approximation that 
$H^{(1-q) \, {\mathfrak{D}}-1} \approx H^{(1-q) \, {\mathfrak{D}}}$.
Rewriting the equation (\ref{temp_def_ie}) based on the assumption yields the expression 
\beq
\exp\left( \frac{1-q^{\prime}}{1-q} \right) \frac{1}{{\mathfrak{D}} \, k \, T} = 
\Xi_{ie}^{1-q} \, H^{(1-q) \, {\mathfrak{D}}} \, 
\exp\left(\frac{1-q^{\prime}}{1-q} \, (\Xi_{ie} \; {H^{\mathfrak{D}}})^{1-q} \right).
\label{mve_lambw_func}
\eeq
The solution of the above equation yields the enthalpy in terms of the Lambert's $W$-function
\beq
H = \left[\frac{1-q}{1-q^{\prime}} \; \frac{1}{\Xi_{ie}^{1-q}} \; 
    W_{0} \left(\frac{1-q^{\prime}}{1-q} \, \exp\left( \frac{1-q^{\prime}}{1-q}\right) 
    \frac{\beta}{{\mathfrak{D}}} \right)\right]^
    {\frac{1}{(1-q) {\mathfrak{D}}}}.
\label{enth_expr_ie}
\eeq
The requirement that the entropy should be concave, along with the fact that the enthalpy is a
continuous function restricts our choice of the $W$ function to the principal branch. 
The specific heat at constant pressure computed from the enthalpy (\ref{enth_expr_ie}) is
\beq
C_{P} \equiv \frac{\partial H}{\partial T}= - \frac{k \, \beta}{(1-q) \, {\mathfrak{D}}} 
                                                \; \frac{W_{0}(\mathfrak{b} \, \beta)}{1 + W_{0} (\mathfrak{b} \, \beta)} \;
          \left[\mathfrak{a}\,  W_{0} (\mathfrak{b} \, \beta)\right]^{\frac{1-(1-q){\mathfrak{D}}}{(1-q) {\mathfrak{D}}}},
\label{spht_ie_expr}
\eeq
where, the factors $\mathfrak{a}$ and $\mathfrak{b}$ are as defined below:
\beq
{\mathfrak{a}} = \frac{1-q}{1-q^{\prime}} \; \frac{1}{\Xi_{ie}^{1-q}}, \qquad
{\mathfrak{b}} = \frac{1-q^{\prime}}{1-q} \, \exp\left( \frac{1-q^{\prime}}{1-q}\right) 
                 \frac{1}{{\mathfrak{D}}}.
\label{mfab_def}
\eeq
In the large $N$ limit the expression for the heat capacity reads:
\beq
C_{P} = - \frac{k \, \beta}{(1-q) \, {\mathfrak{D}}} \; \frac{1}{{\mathfrak{a}}(1 + W_{0} (\mathfrak{b} \, \beta))}.
\label{spht_lgN}
\eeq
The specific heat is positive in the regions (a) and (c) and negative in the region (b).  Since we adopted 
the large $N$ limit in evaluating the enthalpy (\ref{enth_expr_ie}) we do not recover the corresponding Boltzmann Gibbs result, 
because the extensive limit and the large $N$ limit do not commute with each other. 
\subsection{$(\mu,V,\mathsf{L})$ ensemble}
The Hill energy $\mathsf{L}$ is the heat function corresponding to the $(\mu,V,\mathsf{L})$ ensemble. The 
phase space volume enclosed by the curve of constant Hill energy $\mathsf{L}$ is 
\beq
\varSigma(\mu,V,\mathsf{L}) = \sum_{} \frac{1}{N! \, h^{DN}} \; \int_{r_{i}} \int_{p_{i}}
                   \Theta\left(\sum_{i} \frac{p_{i}^{2}}{2m} -\mu N - \mathsf{L}\right)\;
                   \prod_{i=1}^{N} {\rm d}^{D} r_{i}\, {\rm d}^{D} p_{i}.
\label{nad_psv_ig}
\eeq
Integrating over the phase space variables $p_{i}$ and $r_{i}$, we arrive at
\beq
\varSigma(\mu,V,\mathsf{L}) = \sum_{N=0}^{\infty} \frac{V^{N}}{N!} \; \frac{\mathcal{M}^{N}}{\Gamma\left(\frac{D N}{2}+1\right)} \;
                                (\mathsf{L} + \mu N)^{\frac{DN}{2}}.
\label{nad_psv_nsum}
\eeq
An exact evaluation of the summation given in (\ref{nad_psv_nsum}) could not be achieved. In order to calculate an approximate value
in the large $N$ limit we carry out the following procedure. First we approximate the factor $(\mathsf{L} + \mu N)^{\frac{DN}{2}}$
as follows
\bea
(\mathsf{L} + \mu N)^{\frac{DN}{2}} &=& (\mu N)^{\frac{DN}{2}} \, \left(1 + \frac{\mathsf{L}}{\mu N}\right)^{\frac{DN}{2}}, \nn \\
				    &=& (\mu N)^{\frac{DN}{2}} \, \exp\left(\frac{DN}{2}\ln\left(1 + \frac{\mathsf{L}}{\mu N}\right)\right).
\label{npn_approx}
\eea
For very small values of $x$ we can make use of the approximation $\ln(1+x) \approx x$ and this leads to
\beq
(\mathsf{L} + \mu N)^{\frac{DN}{2}} = (\mu N)^{\frac{DN}{2}} \exp\left(\frac{D}{2} \, \frac{\mathsf{L}}{\mu}\right).
\label{npn_approx_fe}
\eeq
In the next step we use the Stirling's approximation for the gamma function as given below:
\beq
\Gamma\left(\frac{DN}{2} +1\right) \approx \left(\frac{DN}{2}\right)^{\frac{DN}{2}} \exp\left(- \frac{DN}{2}\right).
\label{str_approx}
\eeq
Substituting the relations (\ref{npn_approx_fe}), and (\ref{str_approx}) in (\ref{nad_psv_nsum}) we arrive at
\beq
\varSigma(\mu,V,\mathsf{L}) = \exp\left(\frac{D}{2} \, \frac{\mathsf{L}}{\mu}\right) \; \sum_{N=0}^{\infty}\;  \frac{V^{N}}{N!} \; 
                              \frac{\mathcal{M}^{N}}{\left(\frac{DN}{2}\right)^{\frac{DN}{2}} \exp\left(- \frac{DN}{2}\right)} \;
                              (\mu N)^{\frac{DN}{2}}.
\label{psv_sum_expr}
\eeq
Now the summation over $N$ can be carried out in (\ref{psv_sum_expr}), enabling us to write the approximate 
expression for the phase space volume
\beq
\varSigma(\mu,V,\mathsf{L}) = \exp\left(\frac{D}{2} \, \frac{\mathsf{L}}{\mu}\right) \;  
                             \exp\left(\frac{V \, \mu^{D/2} \, e^{D/2} \, \mathcal{M}}{(D/2)^{D/2}}\right).
\label{psv_4e_fe}
\eeq
The approximate computation used by us to calculate the phase space volume (\ref{psv_4e_fe}) can be 
considered  as a first order approximation of the summation in (\ref{nad_psv_nsum}). 
Similarly the surface area enclosed by the curve of constant ${\mathsf{L}}$ is also found
\beq
\Omega(\mu,V,\mathsf{L}) = \frac{D}{2 \mu} \; \exp\left(\frac{D}{2} \, \frac{\mathsf{L}}{\mu}\right) \;  
                             \exp\left(\frac{V \, \mu^{D/2} \, e^{D/2} \, \mathcal{M}}{(D/2)^{D/2}}\right).
\label{psa_4e_fe}
\eeq
The entropy of the classical ideal gas in this adiabatic ensemble found from (\ref{psv_4e_fe}) is 
\beq
S_{q,q^{\prime}} = \frac{k}{1-q^{\prime}} \; \left[\exp \left(\frac{1-q^{\prime}}{1-q} 
                   \left(\left(\Xi_{he} \; \exp\left(\frac{D}{2} \, \frac{\mathsf{L}}{\mu}\right)\right)^{1-q} -1\right)\right) -1\right], 
\eeq
where, for the sake of convenience we define:
\beq
\Xi_{he} = \exp\left(\frac{V \, \mu^{D/2} \, e^{D/2} \, \mathcal{M}}{(D/2)^{D/2}}\right).
\eeq
The temperature of this ensemble calculated from the defining relation (\ref{temp_def_TS}) is 
\beq
\frac{1}{T} = \frac{D}{2 \mu} \, k \, \Xi_{he}^{1-q} \,  \left(\exp\left(\frac{D}{2} \, \frac{\mathsf{L}}{\mu}\right)\right)^{1-q}
		   \exp \left(\frac{1-q^{\prime}}{1-q} 
                   \left(\left(\Xi_{he} \; \exp\left(\frac{D}{2} \, \frac{\mathsf{L}}{\mu}\right)\right)^{1-q} -1\right)\right).
\label{temp_4e_fe} 
\eeq
Rewriting (\ref{temp_4e_fe}) we immediately recognize it as the equation whose solution is given by the 
Lambert's $W$-function. Thus the Hill energy in terms of the $W$-function reads:
\beq
\mathsf{L} =  \frac{2 \, \mu}{(1-q) \, D} \; \ln \left[\frac{1-q}{1-q^{\prime}} \; \frac{1}{\Xi_{he}^{1-q}} \; 
    W_{0}  \left(\frac{1-q^{\prime}}{1-q} \, \exp\left( \frac{1-q^{\prime}}{1-q}\right) 
    \frac{2 \, \beta \mu}{D} \right)\right].
\label{he_ig_fe}
\eeq
The principal branch of the $W$ function is chosen based on the concavity requirement on the entropy
and the continuity of the Hill energy. The Hill energy $\mathsf{L}$ is uniformly positive in the region
(c) for which $q^{\prime} > 1$ and $-\infty < q < 1$, whereas in the other two regions (b) and (c) it is positive 
only when the argument of the logarithm is less than $1$. From (\ref{he_ig_fe}) the specific heat at constant 
volume is found
\beq
C_{V} \equiv \frac{\partial \mathsf{L}}{\partial T} = - \frac{\mu}{(1-q) D/2} \, \frac{k \beta}{1 + W_{0} (\mathtt{b} \, \beta)}, \qquad
\mathtt{b} = \frac{1-q^{\prime}}{1-q} \, \exp\left(\frac{1-q^{\prime}}{1-q}\right) \, \frac{\mu}{D/2}.
\label{sph_he_fe}
\eeq
Subject to the condition $q + q^{\prime} > 1$ and excluding the region $0 \leq (q,q^{\prime}) \leq 1$, 
the specific heat is positive in the region $q > 1$ and negative in the region $q < 1$. Due to the noncommutative 
nature of the extensive limit and the large $N$ limit, we are unable to recover the Boltzmann Gibbs result 
corresponding to the Hill energy (\ref{he_ig_fe}) and the heat capacity (\ref{sph_he_fe}). 
\subsection{$(\mu,P,\mathsf{R})$ ensemble}
The adiabatic ensemble with both the number and volume fluctuations is illustrated using the classical ideal gas
in this section.  The phase space volume of the classical ideal gas in this ensemble is 
\beq
\varSigma(\mu,P,\mathsf{R}) = \sum_{N} \sum_{V} \frac{1}{N! \, h^{DN}} \; \int_{r_{i}} \int_{p_{i}}
                   \Theta\left(\sum_{i} \frac{p_{i}^{2}}{2m} + PV -\mu N - \mathsf{R}\right)\;
                   \prod_{i=1}^{N} {\rm d}^{D} r_{i}\, {\rm d}^{D} p_{i}.
\label{psv_re_ig}
\eeq
Integrating over the phase space co-ordinates namely $p_{i}$ and $r_{i}$ we arrive at
\beq
\varSigma(\mu,P,\mathsf{R}) = \sum_{N=0}^{\infty} \sum_{V} \frac{V^{N}}{N!} \; \frac{\mathcal{M}^{N}}{\Gamma\left(\frac{D N}{2}+1\right)} \;
                                (\mathsf{R} + \mu N - PV)^{\frac{DN}{2}}.
\label{psv_re_vnsum}
\eeq
The sum over volume eigenstates is calculated by approximating it to an integral. Since a direct integration
will lead to over counting of the volume states, we use the shell particle method of counting 
which was developed in [\cite{DC1},\cite{DC2}]. Using this technique eliminates the redundant volume states
and the obtained expression for the phase space volume reads:
\beq
\varSigma(\mu,P,\mathsf{R}) = \sum_{N=0}^{\infty} \mathcal{M}^{N} \;
               \left(\frac{1}{P}\right)^{N} \;
               \frac{(\mathsf{R} + \mu N)^{\mathfrak{D}} }
               { \Gamma \left(\mathfrak{D} + 1 \right) }. 
\label{psv_nsum_re}
\eeq
Finally we evaluate the summation over the number of particles in (\ref{psv_nsum_re}) using a large $N$ approximation. 
The large $N$ limit of the $(\mathsf{R} + \mu N)^{\mathfrak{D}}$ factor is calculated as follows
\beq
(\mathsf{R} + \mu N)^{\mathfrak{D}} = (\mu N)^{\mathfrak{D}} \, \left(1 + \frac{\mathsf{R}}{\mu N}\right)^{\mathfrak{D}}
                                         = (\mu N)^{\mathfrak{D}} \, \exp\left(\mathfrak{D}
                                            \ln\left(1 + \frac{\mathsf{R}}{\mu N}\right)\right).
\label{Nbinom_approx}
\eeq
Using the approximation $\ln(1+x) \approx x$ for small values of $x$ we get
\beq
(\mathsf{R} + \mu N)^{\mathfrak{D}} = (\mu N)^{\mathfrak{D}} \, \exp\left(\frac{\mathcal{D} \, \mathsf{R}}{\mu}\right).
\label{Nbinom_approx_fe}
\eeq
Based on the Stirling's approximation the Gamma function in (\ref{psv_nsum_re}) is written as 
\beq
\Gamma\left(\mathfrak{D} + 1\right) \approx \mathfrak{D}^{\mathfrak{D}} \, \exp\left(-\mathfrak{D}\right).
\label{gf_str_re}
\eeq
Substituting (\ref{Nbinom_approx_fe}) and (\ref{gf_str_re}), we can rewrite (\ref{psv_nsum_re}) as follows
\beq
\varSigma(\mu,P,\mathsf{R}) = \exp\left(\frac{\mathcal{D} \, \mathsf{R}}{\mu}\right) \, \sum_{N=0}^{\infty}\,
                               \left(\frac{\mathcal{M}}{P}\right)^{N} \left(\frac{\mu}{\mathcal{D}}\right)^{\mathfrak{D}} 
                               \exp\left(\mathfrak{D}\right),
\label{psv_nsum_fe}
\eeq
where $\mathcal{D} = \mathfrak{D}/N$.
Carrying out the summation in (\ref{psv_nsum_fe}), we get the final expression of the phase space volume 
in the large $N$ limit
\beq
\varSigma(\mu,P,\mathsf{R}) = \exp\left(\frac{\mathcal{D} \, \mathsf{R}}{\mu}\right)
                              \left[ 1 - \frac{\mathcal{M}}{P} \, \left(\frac{\mu}{\mathcal{D}}\right)^{\mathcal{D}} \,
                              \exp\left(\mathcal{D}\right) \right]^{-1}.
\label{psv_fe_LNL}
\eeq
The computed phase space volume (\ref{psv_fe_LNL}) can be considered as a first order approximation of
(\ref{psv_nsum_re}). The surface area of the phase space curve is 
\beq
\Omega(\mu,P,\mathsf{R}) = \frac{\mathcal{D}}{\mu} \;
                           \exp\left(\frac{\mathcal{D} \, \mathsf{R}}{\mu}\right)
                           \left[1 - \frac{\mathcal{M}}{P} \left(\frac{\mu}{\mathcal{D}}\right)^{\mathcal{D}} \,
                           \exp\left(\mathcal{D}\right) \right]^{-1}.                             
\eeq
The two parameter entropy of the classical ideal gas in the $(\mu,P,\mathsf{R})$ ensemble can be immediately obtained from 
the phase space volume (\ref{psv_fe_LNL}) and reads: 
\beq
S_{q,q^{\prime}} = \frac{k}{1-q^{\prime}} \; \left[\exp \left(\frac{1-q^{\prime}}{1-q} 
                   \left(\left(\Xi_{re} \; \exp\left( 
                   \frac{\mathcal{D} \, \mathsf{R}}{\mu}\right)\right)^{1-q} -1\right)\right) -1\right], 
\label{entr_re_ig}
\eeq
where the temperature independent factor $\Xi_{re}$ of (\ref{entr_re_ig}) is
\beq
\Xi_{re} = \left[1 - \frac{\mathcal{M}}{P} \, \left(\frac{\mu}{\mathcal{D}}\right)^{\mathcal{D}} \,
                              \exp\left(\mathcal{D}\right) \right]^{-1}.
\label{xi_re_def}
\eeq
The temperature of the $(\mu,P,\mathsf{R})$ ensemble is 
\beq
\frac{1}{T}\equiv \frac{\partial S_{q,q^{\prime}}}{\partial \mathsf{R}} 
                  =  \frac{\mathcal{D}}{\mu} \, \Xi_{re}^{1-q} \,  
                      \left(\exp\left(\frac{ \mathcal{D} \, \mathsf{R}}{\mu}\right)\right)^{1-q}            
		     \exp \left(\frac{1-q^{\prime}}{1-q} 
                     \left(\left(\Xi_{re} \; \exp\left( \frac{\mathcal{D} \, \mathsf{R}}{\mu}\right) \right)^{1-q} -1\right)\right).
\label{temp_re_def} 
\eeq
Inverting the expression for the temperature in (\ref{temp_re_def}) helps us in computing the Ray energy,  
the heat function corresponding to the $(\mu,P,\mathsf{R})$ ensemble.  The expression for the Ray energy of the
classical ideal gas reads: 
\beq
\mathsf{R} =  \frac{\mu}{(1-q) \, \mathcal{D}} \; \ln \left[\frac{1-q}{1-q^{\prime}} \; \frac{1}{\Xi_{re}^{1-q}} \; 
    W_{0} \left(\frac{1-q^{\prime}}{1-q} \, \exp\left( \frac{1-q^{\prime}}{1-q}\right) 
    \frac{ \beta \mu}{\mathcal{D}} \right)\right].
\label{re_ig_fe}
\eeq
The continuity requirement of the Ray energy $\mathsf{R}$, and the concavity conditions on the entropy
restricts our choice of the $W$-function to the principal branch. In the region (c) for which 
$q^{\prime} > 1$ and $-\infty < q < 1$ the Ray energy is uniformly positive whereas in the other 
two regions {\it viz} (a) and (b) it is positive only when the argument of the logarithm is less than
$1$. From (\ref{re_ig_fe}) the specific heat at constant pressure is found: 
\beq
C_{P} \equiv \frac{\partial R}{\partial T} = - \frac{\mu}{(1-q) \, \mathcal{D}} \, \frac{k\, \beta} {1 + W_{0} (\mathbf{b} \, \beta)},
\qquad
\mathbf{b} = \frac{1-q^{\prime}}{1-q} \, \exp\left(\frac{1-q^{\prime}}{1-q} \right) \, \frac{\mu}{\mathcal{D}}.
\label{spht_re}
\eeq
Analogous to the $(\mu,V,\mathsf{L})$ the specific heat is positive in the region $q < 1$ and negative in the
region $q > 1$ under the condition that $q + q^{\prime} > 1$ and excluding the region $0 \leq (q,q^{\prime}) \leq 1$. 
Further the Boltzmann Gibbs results could not be recovered from (\ref{re_ig_fe}) and (\ref{spht_re}) since we have used 
the large $N$ limit in the computation of the density of states and it does not commute with the extensive 
$(q,q^{\prime}) \rightarrow 1$ limit.
%
%
%
%
%
%
\setcounter{equation}{0}
\section{Gas molecules under gravity}
\label{gravity}
In order to investigate the effect of an external field on a thermodynamic
system we study a system of gas molecules in the presence of a gravitational field in 
the current section. A system of $D$-dimensional classical ideal gas under the effect 
of gravity is explored in the first part of the section.  To understand an interacting 
system in an external field, we, in the later part of the section, 
examine a system of hard rods confined in a linear box of length $L$ in the 
presence of an external gravitational field.
\subsection{D-dimensional ideal gas under gravity}
A system of $N$ noninteracting classical point-like particles of mass $m$ is considered
in $D$-dimensions in the presence of a uniform gravitational field $g$. The acceleration 
due to gravity acts along one of the dimensions labelled
as the $z$ co-ordinate and the force is oriented in a direction opposite to that of the 
co-ordinate axis.  To conform to reality we consider that the gas molecules are located 
within a fixed region in space and so the position co-ordinates take values over a finite
interval. Of the $D$ position co-ordinates denoted by $r_{i} \; \; (i= 1,2,...,D)$, 
the $z$ co-ordinate ranges over the interval $(0,\mathsf{H})$ and the remaining 
$D-1$ co-ordinates range over the values $(0,L)$. The Hamiltonian of the $D$-dimensional 
ideal gas under gravity is
\beq
\mathcal{H} = \sum_{i=1}^{DN} \frac{p_{i}^{2}}{2m} + \sum_{j = 1}^{N} mgz_{j} + U.
\label{ham_ug_D}
\eeq
The fact that the gas molecules are to exist within a finite region in space is indicated 
in the Hamiltonian through a potential which mimics the presence of a wall
\bea
U =  \begin{cases} 
               0 & {\text{within the allowed region}}, \\
               \infty & {\text{outside the allowed region}}.
               \end{cases}
\label{pot_cond}
\eea
Substituting the expression for the Hamiltonian (\ref{ham_ug_D}) in the integral for
the phase space volume enclosed by a constant energy curve, we arrive at 
\beq
\varSigma(N,V,E) =  \frac{1}{N! \, h^{DN}} \; \int_{r_{i}} \int_{p_{i}}
                   \Theta\left(\sum_{i=1}^{DN} \frac{p_{i}^{2}}{2m} 
                   + \sum_{j=1}^{N} mgz_{j} + U - E\right)\;
                   \prod_{i=1}^{N} {\rm d}^{D} r_{i}\, {\rm d}^{D} p_{i}.
\label{psv_int_grg}
\eeq
Integrating over the phase space co-ordinates, the volume enclosed by the constant energy curve is 
\beq
\varSigma(N,V,E) = \frac{V^{N}}{N! \, h^{DN}} \, \frac{\mathcal{M}^{N}}
                   {\Gamma\left(\mathfrak{D}+1\right) \, (mg\mathsf{H})^{N}} \
                   \mathfrak{S}_{D} \left(E,mg \mathsf{H}\right),
\label{psv_fe_grg}
\eeq
where $\mathfrak{S}(a,b)$ is a finite sum defined as below:
\beq
\mathfrak{S}_{D} \left(a,b \right) = 
                   \sum_{k=0}^{N} (-1)^{k} \,  \binom{N}{k} \, (a - kb)^{\mathfrak{D}} \, 
                   \Theta(a - k b).
\label{sum_def}
\eeq
The surface area of the constant energy curve in the phase space is 
\beq
\Omega(N,V,E) =  \frac{V^{N}}{N! \, h^{DN}} \, \frac{\mathcal{M}^{N}}{\Gamma\left(\mathfrak{D}\right) \, (mg\mathsf{H})^{N}} \
                 \mathfrak{S}^{\prime}_{D} \left(E,mg \mathsf{H}\right),
\label{psa_fe_grg}
\eeq
where the prime over the summation (\ref{sum_def}) denotes a partial derivative with respect to the energy.
Substituting the phase space volume (\ref{psv_fe_grg}) in equation (\ref{entr_def_rel}) we obtain the entropy of
a $D$-dimensional ideal gas in the presence of gravity
\beq
S_{q,q^{\prime}} = \frac{k}{1-q^{\prime}} \; \left[\exp \left(\frac{1-q^{\prime}}{1-q} 
                   \left(\left(\Xi_{g} \; \mathfrak{S}_{D} \left(E,mg \mathsf{H}\right)
                   \right)^{1-q} -1\right)\right) -1\right],
\label{entr_grg_fe} 
\eeq
where, the factor $\Xi_{g}$ is defined as
\beq
\Xi_{g} = \frac{A^{N}}{N! \, h^{DN}} \, \frac{\mathcal{M}^{N}}{\Gamma\left(\mathfrak{D}+1\right) \, (mg)^{N}},
\qquad
A = \prod_{\alpha=1}^{D-1} L_{\alpha}.
\label{xi_def}
\eeq
Using the entropic expression (\ref{entr_grg_fe}), and from the 
definition of temperature (\ref{temp_def_TS}) we get 
\bea
\frac{1}{T} =  k \; \Xi_{g}^{1-q} \,
                     \frac{\mathfrak{S}^{\prime}_{D} \left(E,mg \mathsf{H}\right)}
                     {\left(\mathfrak{S}_{D} \left(E,mg \mathsf{H}\right)\right)^{q}} \;
                   \exp \left(\frac{1-q^{\prime}}{1-q} 
                     \left(\left(\Xi_{g} \; \mathfrak{S}_{D} \left(E,mg \mathsf{H}\right) \right)^{1-q} -1\right)\right).
\label{temp_fe_grg}
\eea
Looking at the structure of the relation (\ref{temp_fe_grg}) given above, we realize that an inversion
to obtain the internal energy as a function of temperature is not feasible.  But the specific heat of 
the system can also be computed from the knowledge of the phase space volume (\ref{psv_fe_grg}) and the 
surface area (\ref{psa_fe_grg}) and the temperature (\ref{temp_fe_grg}) via the relation:
\beq
C_{V} \equiv \left( \frac{\partial T}{\partial E} \right)^{-1} 
        = \left[ T \, \left(\frac{q \ \Omega}{\varSigma} - \frac{1}{\Omega} \; \frac{\partial \Omega}{\partial E}
           - (1-q^{\prime}) \frac{\Omega}{\varSigma^{q}}\right) \right]^{-1}.
\label{spht_psv_psa_temp}
\eeq
Substituting the phase space volume (\ref{psv_fe_grg}), area of the curve (\ref{psa_fe_grg}) and 
the expression for the temperature (\ref{temp_fe_grg}) in (\ref{spht_psv_psa_temp}) the specific heat 
as a function of energy reads: 
\bea
C_{V} &=& \mathfrak{D} \, k\, \exp \left(\frac{1-q^{\prime}}{1-q} 
                     \left(\left(\Xi_{g} \; \mathfrak{S}_{D} \left(E,mg \mathsf{H}\right) \right)^{1-q} -1\right)\right) \,
          \bigg[\mathfrak{D} \, q \; (\Xi_{g} \, \mathfrak{S}_{D} \left(E,mg \mathsf{H}\right))^{q-1} \nn \\
      & &  - \Xi_{g}^{q-1} \, \frac{(\mathfrak{S}_{D}  \left(E,mg \mathsf{H}\right))^{q}}
             {\mathfrak{S}_{D}^{\prime} \left(E,mg \mathsf{H}\right)} \,
           \frac{\partial}{\partial E} \ln \mathfrak{S}_{D}^{\prime} \left(E,mg \mathsf{H}\right)
           - (1-q^{\prime}) \, \mathfrak{D} \bigg]^{-1}.
\label{spht_ie_fe}
\eea
From (\ref{spht_ie_fe}) it can be noticed that the specific heat can have both positive and negative values 
depending on the term in square bracket. 
We investigate the relevant limiting cases: {\it(i)} First is the $g \rightarrow 0$ limit in which 
the gas molecules behave like an ideal gas.  {\it(ii)} The second case is the $\mathsf{H} \rightarrow \infty$
limit in which eqn. (\ref{temp_fe_grg}) can be inverted in the large $N$ limit to obtain the energy as a 
function of temperature.  We present the relevant calculation of this limiting case in the discussion below:

In the infinite height limit the limiting value of the summation (\ref{sum_def}) and its derivative are
\beq
\lim_{\mathsf{H} \rightarrow \infty}  
\;  \mathfrak{S}_{D} \left(E,mg \mathsf{H}\right) = E^{\mathfrak{D}}, 
\qquad
\lim_{\mathsf{H} \rightarrow \infty}
\;  \mathfrak{S}_{D}^{\prime} \left(E,mg \mathsf{H}\right) = \mathfrak{D} \, E^{\mathfrak{D}-1}.
\label{sum_lt_ht}
\eeq
Substitution of the limiting value (\ref{sum_lt_ht}) in the expression for the temperature
(\ref{temp_fe_grg}) leads to: 
\beq
\frac{1}{k \, T} = \mathfrak{D} \, k \; \Xi_{g}^{1-q} \, E^{(1-q)\, \mathfrak{D}  - 1} \,
                   \exp \left(\frac{1-q^{\prime}}{1-q} 
                   \big((\Xi_{g} \; E^{\mathfrak{D}})^{1-q} -1\big)\right).
\label{temp_rel_ig}
\eeq
In the large $N$ limit using the approximation 
$E^{(1-q) \,\mathfrak{D} - 1} \approx E^{(1-q) \, \mathfrak{D}}$ 
in (\ref{temp_rel_ig}), we can invert it to obtain the 
internal energy as a function of temperature in terms of the Lambert's W-function
\beq
E = \left[\frac{1-q}{1-q^{\prime}} \; \frac{1}{\Xi_{g}^{1-q}} \; 
    W_{0} \left(\frac{1-q^{\prime}}{1-q} \, \exp\left( \frac{1-q^{\prime}}{1-q}\right) 
    \frac{\beta}{\mathfrak{D}} \right)\right]^
    {\frac{1}{(1-q)\, \mathfrak{D}}}.
\label{IE_rel_ig}
\eeq
The preconditions that the entropy should be concave, and the energy should be a continuous function
helps us to conclude that only the principal branch of the $W$ function occurs in (\ref{IE_rel_ig}). 
The specific heat of the classical ideal gas under gravity computed from the 
internal energy is 
\beq
C_{V} = - \frac{k \, \beta}{(1-q) \, \mathfrak{D}} \; 
          \frac{W_{0} (\mathfrak{b} \, \beta)}{1 + W_{0} (\mathfrak{b} \, \beta)} \;
          \left[\mathsf{a}_{g} \,  W_{0} (\mathfrak{b} \, \beta)\right]^{\frac{1-(1-q)\mathfrak{D}}{(1-q)\mathfrak{D}}},
\label{spht_fe_ig}
\eeq
where the factor $\mathfrak{b}$ was introduced in (\ref{mfab_def}) and $\mathsf{a}_{g}$ is 
\beq
{\mathsf{a}}_{g} = \frac{1-q}{1-q^{\prime}} \, \frac{1}{\Xi_{g}^{1-q}}.
\label{a_def}
\eeq
The large $N$ limit of the specific heat for the ideal gas under gravity reads: 
\beq
C_{V} = - \frac{k \beta}{(1-q) \mathfrak{D}} \; \frac{1}{{\mathsf{a}}_{g} (1 + W_{0} (\mathfrak{b} \beta))}.
\label{spht_lgN_ig}
\eeq
Investigating the expression for the specific heat, we find that it can be either positive in the
regions (a) and (c) or negative in the region (b).  Since we have already made use of the large $N$
limit we do not recover the Boltzmann Gibbs results because the thermodynamic limit and the 
extensive limit do not commute with each other.

The one parameter limits corresponding to the temperature relation (\ref{temp_fe_grg})
can be obtained by allowing either $q$ or $q^{\prime}$ to take the limiting value. We notice
that the expression for the temperature is the same when we set either of the parameter
to unity and reads: 
\beq
\frac{1}{T} =  k \, \Xi_{g}^{1-q} \, 
              \frac{\mathfrak{S}_{D} \left(E,mg \mathsf{H}\right)}
              {\mathfrak{S}_{D}^{\prime} \left(E,mg \mathsf{H}\right)}.
\label{tmp_rel_1p}
\eeq
The definition of temperature (\ref{tmp_rel_1p}) can be inverted to obtain the internal energy 
and the specific heat in the $\mathsf{H} \rightarrow \infty$ limit, but there is no need to invoke 
the large $N$ limit.  
\subsection{1D hard rod gas under gravity}
The generalized statistical mechanics based on the two parameter $(q,q^{\prime})$ entropy 
is applied to a system comprising of $N$ one dimensional hard rods under gravity. 
Initially we consider a system of hard rods of mass $m$ and length $\sigma$ in a finite
region of the space under the influence of a uniform gravitational field of strength $g$. 
The position of the centers and the momentum of the hard rods are denoted by
the set of values $(z_{1},...,z_{N};p_{1},...,p_{N})$. Two rods with centers given by 
$z_{0}$ and $z_{N+1}$ are considered so that their edges define the boundaries of the region, in 
such way that $z_{N+1} - z_{0} = L + \sigma$. 
The relevant thermodynamic quantities like the entropy, the temperature, and 
the heat capacity are then computed.  Later we assume the $L \rightarrow \infty$ limit of the
system and obtain the internal energy in this limiting case. The Hamiltonian of a one 
dimensional hard rod gas under gravity reads:
\beq
\mathcal{H} = \sum_{i=1}^{N} \, \frac{p_{i}^{2}}{2m} + \sum_{i=1}^{N} mgz_{i}
              + \sum_{i<j} u(|z_{j}-z_{i}|) + \sum_{i=1}^{N} U(z_{i};z_{0},z_{N+1}).
\label{ham_hrdg}
\eeq
The distance between the centers of any two particles in the system cannot be less than
the length of the rod $\sigma$ and the potential corresponding to it is 
\bea
u(|z_{j}-z_{i}|) = \begin{cases}
                        \infty, & |z_{j}-z_{i}| < \sigma, \\
                        0, & |z_{j}-z_{i}| \geq \sigma.
                   \end{cases}
\label{pot_int_cond}
\eea
Similarly the existence of the gas molecules within the finite region of the space is 
indicated in the Hamiltonian (\ref{ham_hrdg}) via the potential
\bea
U =  \begin{cases} 
               0, & z_{0} + \sigma \leq z_{i} \leq z_{N+1} - \sigma, \\
               \infty, & {\text{otherwise}}.
               \end{cases}
\label{pot_hrdg_cond}
\eea
Using the Hamiltonian (\ref{ham_hrdg}) in (\ref{vds_en_TS}), the integral expression corresponding to
 the phase space volume is 
\beq
\varSigma(N,L,E) =  \frac{1}{N! \, h^{N}} \; \int_{z_{i}} \int_{p_{i}}
                   \Theta\left(\sum_{i=1}^{N} \frac{p_{i}^{2}}{2m} 
                   + \sum_{j=1}^{N} mgz_{j} + U + u(|z_{j}-z_{i}|)- E\right)\;
                   \prod_{i=1}^{N} {\rm d} z_{i}\, {\rm d} p_{i}.
\label{psv_int_ghrg}
\eeq
To evaluate the phase space volume we first integrate over the momentum variables.  In the next 
step we use the substitution $r_{k} = z_{k} - (z_{0} + k \sigma)$  for the position 
integrals which helps us in integrating over the position variables.  The final expression for the
phase space volume thus computed is 
\beq
\varSigma(N,L,E) =  \frac{\mathcal{M}_{1}^{N}}{N! \; \Gamma\left(\frac{3 N}{2}+1\right)\; (mg)^{N}} \;
                    \mathfrak{S}_{1}(E - mg \alpha,mg \bar{L}).
\label{psv_hrdg_fe}
\eeq
where the $\mathcal{M}_{1}$ is obtained by setting $D=1$ in the expression for $\mathcal{M}$.  For convenience
we have also defined the following quantities
\beq
\alpha =  N z_{0} + \frac{N(N+1)}{2} \, \sigma, \qquad
\bar{L} = L - N \sigma.
\eeq
The surface area of the constant energy curve is 
\beq
\Omega(N,L,E) =  \frac{\mathcal{M}_{1}^{N}}{N! \; \Gamma\left(\frac{3 N}{2}\right) \; (mg)^{N}} \;
                \mathfrak{S}_{1}^{\prime}(E - mg \alpha,mg \bar{L}).
\label{psa_hrdg_fe}
\eeq
The knowledge of the phase space volume (\ref{psv_hrdg_fe}) enables us to obtain the entropy of the 
system defined through the relation (\ref{entr_def_rel}). Combining this with the definition of temperature 
(\ref{temp_def_TS}), we arrive at
\beq
\frac{1}{T} = k\, \Xi_{gh}^{1-q} \, \frac{\mathfrak{S}_{1}^{\prime}(E - mg \alpha,mg \bar{L})}
                  {(\mathfrak{S}_{1}(E - mg \alpha,mg \bar{L}))^{q}} 
               \exp\left(\frac{1-q^{\prime}}{1-q}\left((\Xi_{gh} \,
                            \mathfrak{S}_{1}(E - mg \alpha,mg \bar{L}))^{1-q} - 1 \right)\right),
\label{temp_hrdg_rel}
\eeq
where the factor $\Xi_{gh}$ is defined as
\beq
\Xi_{gh} = \frac{\mathcal{M}_{1}^{N}}{N! \; \Gamma\left(\frac{3 N}{2}+1\right)\; (mg)^{N}}.
\eeq
The form of (\ref{temp_hrdg_rel}) makes us realize that an exact inversion to obtain the internal energy is 
not feasible. To overcome this we use (\ref{spht_psv_psa_temp}) in conjunction with (\ref{psv_hrdg_fe}),
(\ref{psa_hrdg_fe}) and (\ref{temp_hrdg_rel}) to compute the specific heat as a function of the internal energy
and the free length $(\bar{L})$. The specific heat thus evaluated is
\bea
C_{V} &=& \frac{3N}{2} k\, \exp \left(\frac{1-q^{\prime}}{1-q} 
                     \left(\left(\Xi_{gh} \; \mathfrak{S}_{1} \left(E - mg \alpha,mg \bar{L}\right) \right)^{1-q} -1\right)\right) \,
\label{spht_ie_hdg}          \\
      & &  \bigg[\frac{3N}{2} \, q \; (\Xi_{gh} \, \mathfrak{S}_{1} \left(E - mg \alpha,mg \bar{L}\right))^{q-1} \nn \\
      & &    - \Xi_{gh}^{q-1} \, \frac{(\mathfrak{S}_{1}  \left(E - mg \alpha,mg \bar{L}\right))^{q}}
             {\mathfrak{S}_{1}^{\prime} \left(E - mg \alpha,mg \bar{L}\right)} \,
           \frac{\partial}{\partial E} \ln \mathfrak{S}_{1}^{\prime} \left(E - mg \alpha,mg \bar{L}\right)
           - (1-q^{\prime}) \, \frac{3N}{2} \bigg]^{-1}. \nn 
\eea
Analyzing (\ref{spht_ie_hdg}) we realize that both positive and negative values of specific heat are permissible
depending on the argument within the square bracket. There are two pertinent limiting cases namely {\it (i)}
the $g \rightarrow 0$ limit wherein it becomes a system of hard rods moving in a length $L$, and {\it (ii)} the
$L \rightarrow \infty$ limit. In the second limit we present the expression for temperature below:    
\beq
\frac{1}{T} = \frac{3N}{2} k \, \Xi_{gh}^{1-q} \, (E - mg\alpha)^{(1-q)\frac{3N}{2} - 1} \,
               \exp\left(\frac{1-q^{\prime}}{1-q}\left(\left(\Xi_{gh} \; (E - mg\alpha)^{\frac{3N}{2}}\right)^{1-q} - 1 \right) \right),
\label{temp_hrdg_lNlt}
\eeq
which can be inverted in the large $N$ limit by assuming $(3N/2)-1 \approx (3N/2)$. This approximation 
yields the internal energy in terms of the Lambert's $W$-function
\beq
E = \left[{\mathsf{a}}_{gh} \; 
    W_{0} \left(\frac{1-q^{\prime}}{1-q} \, \exp\left( \frac{1-q^{\prime}}{1-q}\right) 
    \frac{2 \beta}{3N} \right)\right]^
    {\frac{2}{(1-q)\, 3N}} + mg \alpha,
\qquad
{\mathsf{a}}_{gh} = \frac{1-q}{1-q^{\prime}} \; \frac{1}{\Xi_{gh}^{1-q}}.
\label{ie_hrdg_rel}
\eeq
The stipulations that the entropy should be concave and the energy should be a continuous function limits 
us to the principal branch of the $W$-function. The specific heat of the hard rod gas in the large $N$ limit
can be obtained from (\ref{ie_hrdg_rel}) and reads:
\beq
C_{V} = - \frac{2 k \beta}{(1-q) 3N} \; \frac{1}{{\mathsf{a}}_{gh} (1 + W_{0} (\mathfrak{b}_{h} \beta))},
\qquad 
{\mathfrak{b}}_{h} =  \frac{1-q^{\prime}}{1-q} \, \exp\left( \frac{1-q^{\prime}}{1-q}\right) \ \frac{2}{3N}.
\label{spht_lgN_hrdg}
\eeq
This has the same regimes of positive and negative specific heat values as the classical ideal gas.  

\begin{figure}[!ht]
\begin{center}
\resizebox{75mm}{!}{\includegraphics{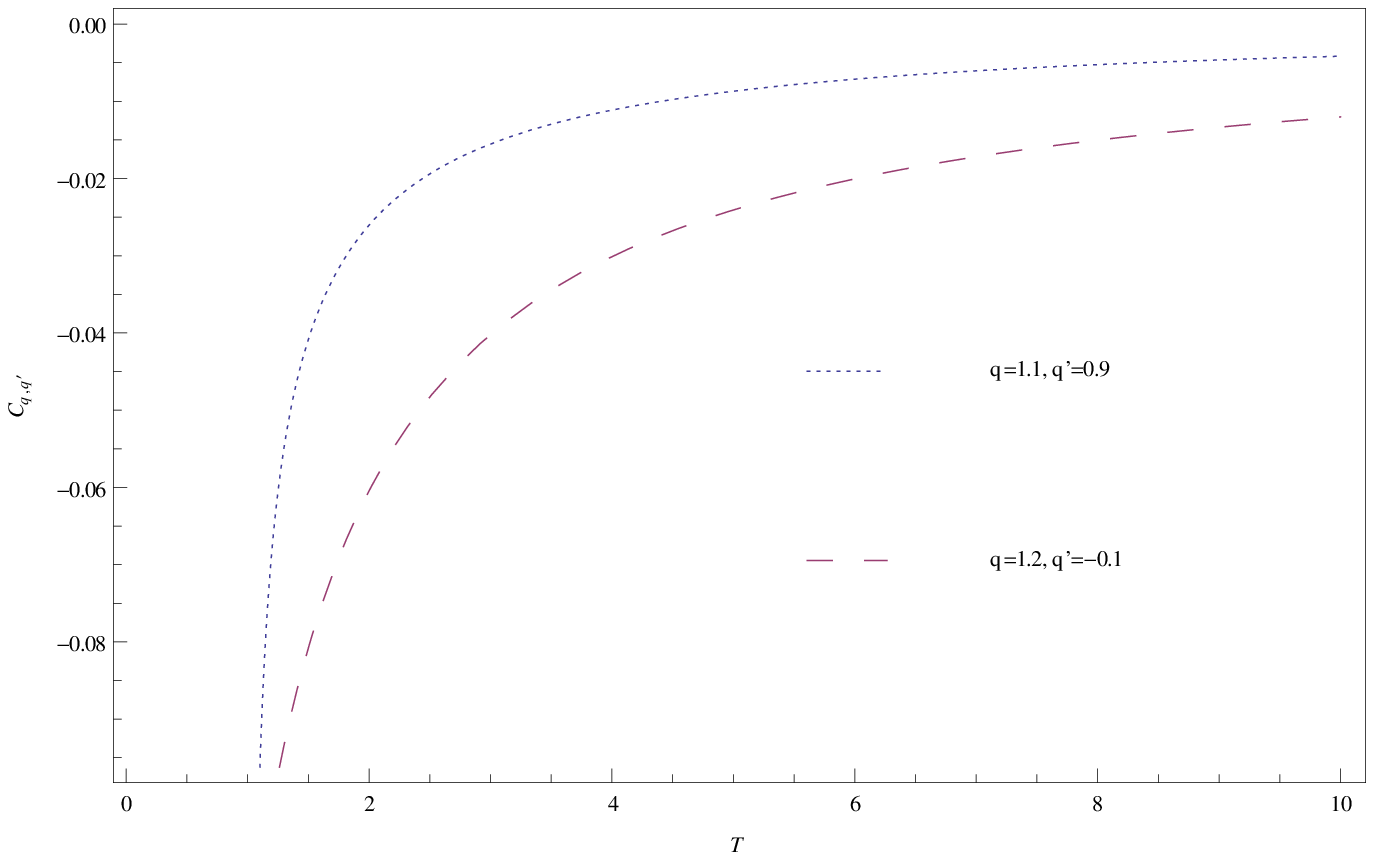}}
\resizebox{75mm}{!}{\includegraphics{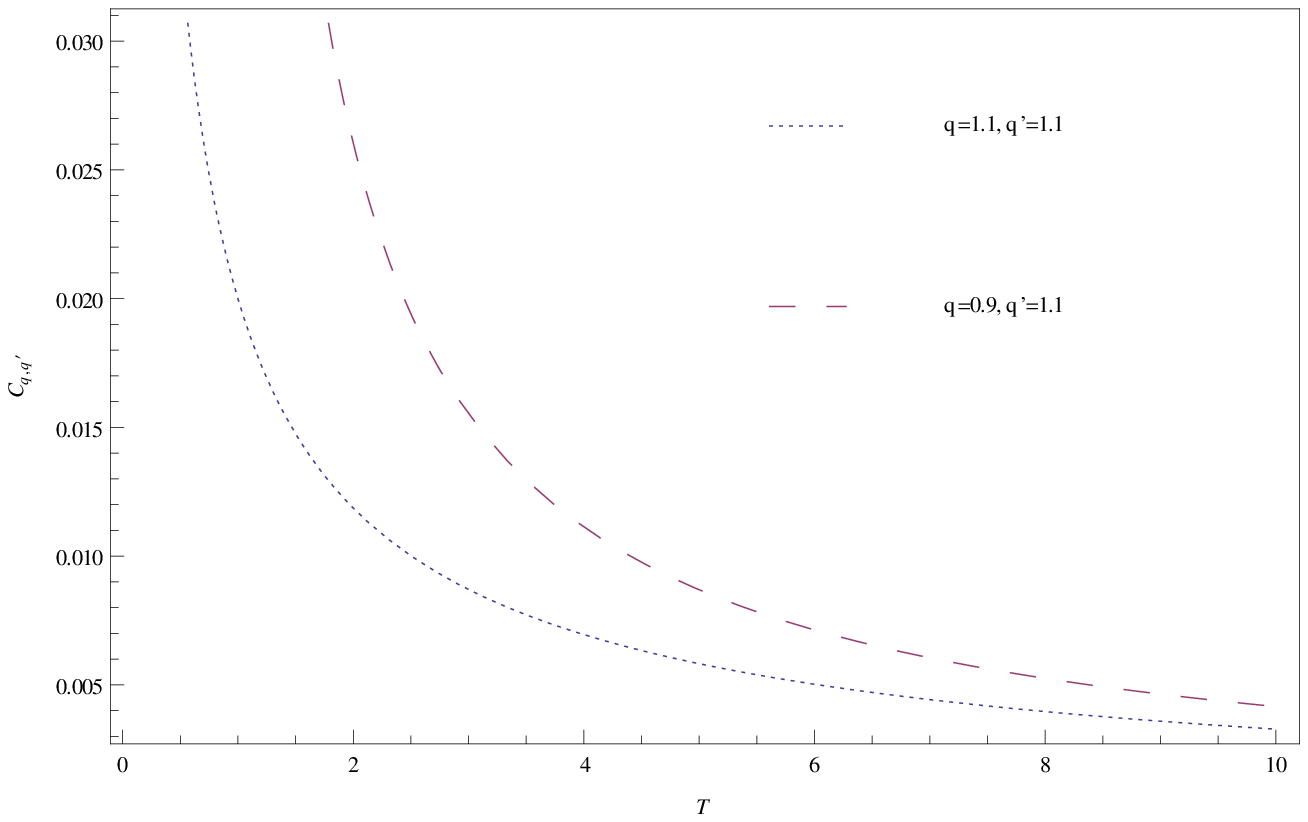}}
\resizebox{75mm}{!}{\includegraphics{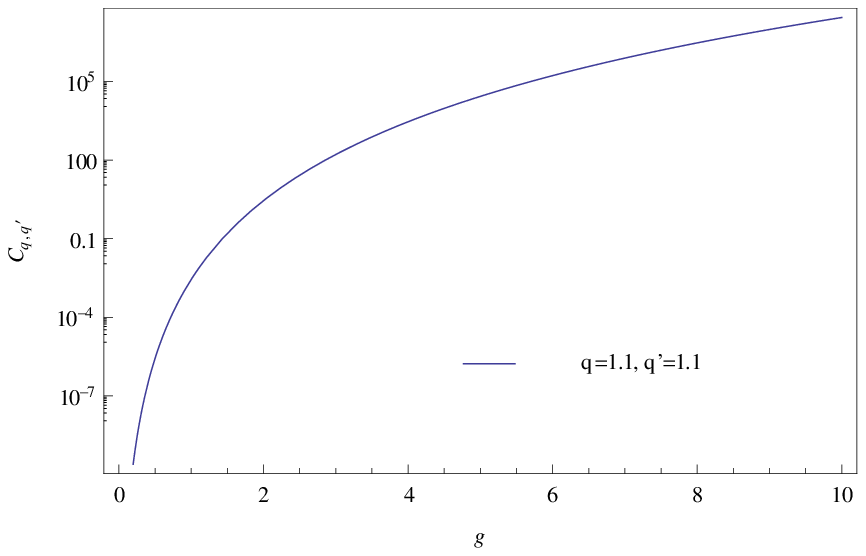}}
\resizebox{75mm}{!}{\includegraphics{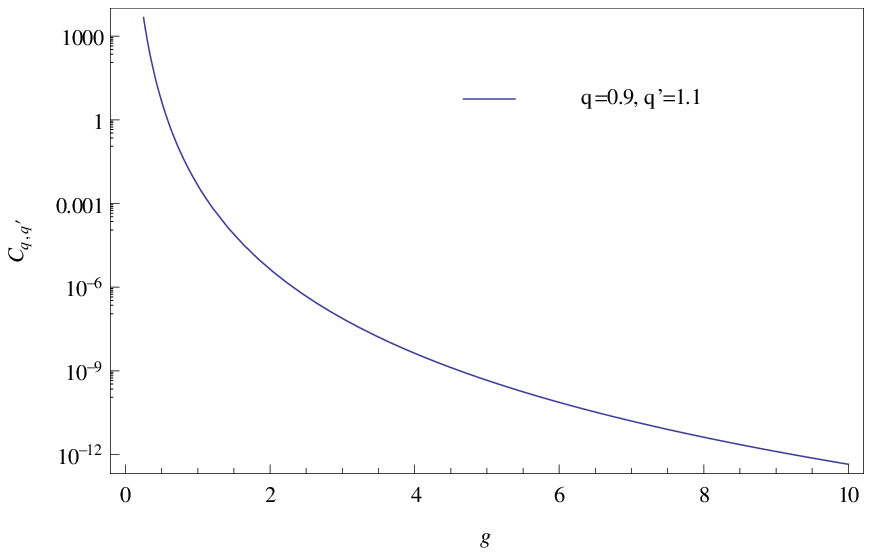}}
\end{center}
\caption{In the following graphs we plot the specific heat (\ref{spht_lgN_hrdg}).  
In the first set of graphs above we plot the variation of the specific heat with respect to the 
temperature at a fixed value of the acceleration due to gravity. The second set of graphs consists 
of the variation of the specific heat with respect to the acceleration due to gravity at a fixed 
temperature. The values of $q$ and $q^{\prime}$ made use of are given in the graphs.}
\end{figure}

The variation of the specific heat with respect to the temperature and the acceleration due to gravity have been 
plotted in the set of graphs above.  Maintaining $g$ at a constant value, we first investigate the variation of 
the specific heat with the temperature. Depending on the value of $q$ and $q^{\prime}$ the specific heat can take 
positive and negative values and show a saturating behaviour at high temperatures. Similarly, we study the 
variation of specific heat with respect to the acceleration due to gravity $g$. The specific heat increases 
(decreases) exponentially when $q$ is greater (lesser) than one.  Previous studies in nonextensive statistical 
mechanics have pointed out that nonextensivity can induce effective interactions [\cite{CCN1},\cite{AL1}] and can also 
describe systems in which collisions generate non-Maxwellian distribution functions [\cite{RGD09}].  A comparison of the 
results obtained in our work, with systems in which a cloud of interacting gas molecules is influenced by 
gravity will help us to appreciate the use of nonextensive statistical mechanics in studying interacting systems.  

The one parameter limits of the temperature (\ref{temp_hrdg_rel}) is investigated.  It can be noticed that 
whether we set $q$ or $q^{\prime}$ to unity is not relevant and we always arrive at 
\beq
\frac{1}{T} = k\, \Xi_{gh}^{1-q} \, \frac{\mathfrak{S}_{1}^{\prime}(E - mg \alpha,mg \bar{L})}
                  {(\mathfrak{S}_{1}(E - mg \alpha,mg \bar{L}))^{q}},
\label{temp_1p_rel}
\eeq 
the temperature relation corresponding to the Tsallis $q$-entropy.  Though the internal energy and the 
specific heat can be obtained from (\ref{temp_1p_rel}) in the $L \rightarrow \infty$ limit, there is no need 
to use the large $N$ limit unlike the two parameter case. 
%
%
%
%
%
\setcounter{equation}{0}
\section{Conclusions}
In the current work we investigate the adiabatic class of ensembles in the framework of generalized statistical mechanics
based on Schw\"{a}mmle-Tsallis $(q,q^{\prime})$ entropy. We do not study the isothermal class of ensembles since the
canonical treatment carried out in [\cite{As2008}] can be extended to the other members of the class which includes
the isothermal-isobaric, the grandcanonical and the generalized ensembles.  We provide a unified description of the 
adiabatic class of ensembles which includes the microcanonical $(N,V,E)$, isoenthalpic-isobaric $(N,P,H)$, the
$(\mu,V,\mathsf{L})$ ensemble, and the $(\mu,P,\mathsf{R})$ ensemble.  A generalized form of the equipartition theorem, 
the virial theorem, and the adiabatic theorem are obtained.  We investigate the nonrelativistic classical ideal gas in 
all the four ensembles. In the microcanonical, and the isoenthalpic-isobaric ensemble, the entropy could be found for 
an arbitrary number of particles. Using the large $N$-limit, the heat functions are obtained in terms of the temperature 
and expressed in terms of the Lambert's $W$-function. From the heat functions, the respective specific heats are evaluated. 
To the best of our knowledge an exact evaluation of the phase space volume in the $(\mu,V,\mathsf{L})$ and the 
$(\mu,P,\mathsf{R})$ ensembles has not been done so far.  We, in our current work assume a large $N$ limit and compute 
the approximate phase space volume.  From the phase space volume the entropy, the heat function and the specific heat of 
the classical ideal gas is found. The heat function and the specific heat are obtained in terms of the Lambert's 
$W$-function without any further approximation. The preconditions that the entropy should be concave and the heat function should be a 
continuous function of the deformation parameter restricts our choice of the $W$-function to the principal branch.  The 
two parameter entropy is concave only in the region where $q+q^{\prime} > 1$ excluding the zone $0 < (q,q^{\prime}) < 1$
where the entropy is not always concave and so we analyze the specific heats only in the above mentioned region.
For the microcanonical and the isoenthalpic-isobaric ensemble the specific heat is positive in the region for which
both $q$ and $q^{\prime}$ are greater than one and in the region where $q^{\prime} >1$ and $-\infty < q < 1$.  But it is
negative in the region $q > 1$ and $-\infty < q^{\prime} < 1$.  In the regime where the entropy is concave, the heat capacities 
of the  $(\mu,V,\mathsf{L})$ and the $(\mu,P,\mathsf{R})$ ensembles are positive when $q > 1$ and negative if $q < 1$. 
The microcanonical specific heat of classical ideal gas and a system of hard rod gas confined in a finite region of 
space and subjected to gravity is also analyzed. The entropy and the specific heat are initially calculated as a 
function of internal energy exactly.  But when we assume the height to be infinite and the number of particles to be 
very large, the internal energy could be obtained as a function of temperature in terms of the Lambert's $W$-function.
Again only the principal branch of the $W$-function contributes to the internal energy and the specific heat.  
In the case of the hard rod gas, the free length $\bar{L}$ and the factor $E-mg \alpha$ plays a role analogous to the 
height and the energy in the classical ideal gas. It has been proved in [\cite{AL1}] that the 
study of free ideal gas in deformed statistical mechanics can describe an interacting gas in the ordinary statistical 
mechanics, and the deformation parameter provides information about interactions.  Also we notice from our current study
that the two parameter entropy allows regions of both positive and negative specific heat unlike the case of the 
Tsallis $q$-entropy in which the large $N$-limit permits only regions of negative heat capacity [\cite{SA99}].
Thus the generalized statistical mechanics based on the two parameter entropy shows more rich features and accommodates
much variations due to the presence of one more deformation parameter.  A further application of this generalized
statistical mechanics to study self gravitating systems which exhibit negative specific heat will be worth pursuing. 
 
Application of the large $N$ limit developed in the current article to study the adiabatic ensembles of other entropies
like the $\kappa$-entropy [\cite{GK2001}] and the two parameter Sharma-Mittal-Taneja entropy [\cite{ST75},\cite{M75}]
may help us in understanding the similarities and differences in the thermostatistical structure of different entropies.
Such an understanding may help us to choose the generalized entropy which describes a given system in an appropriate way. 
The construction of a Laplace transform based on the two parameter exponential may be of great help in establishing
the connection between the isothermal and the adiabatic class of ensembles.  The repeated occurrence of the Lambert's
$W$-function in the field of generalized statistical mechanics [\cite{As2008},\cite{FS04},\cite{FS09},\cite{TO07}] 
points to an important connection between them which is worthy of further investigation. 
%
%
%
%
\section*{Acknowledgements}
The authors would like to thank Professor Ranabir Chakrabarti for discussions. 

%
%
%

\end{document}

%% file: wfunction.tex
\setlength{\unitlength}{0.240900pt}
\ifx\plotpoint\undefined\newsavebox{\plotpoint}\fi
\sbox{\plotpoint}{\rule[-0.200pt]{0.400pt}{0.400pt}}%
\begin{picture}(1500,900)(0,0)
\sbox{\plotpoint}{\rule[-0.200pt]{0.400pt}{0.400pt}}%
\put(252,178){\makebox(0,0)[r]{-3}}
\put(252.0,178.0){\rule[-0.200pt]{4.818pt}{0.400pt}}
\put(252,323){\makebox(0,0)[r]{-2}}
\put(252.0,323.0){\rule[-0.200pt]{4.818pt}{0.400pt}}
\put(252,468){\makebox(0,0)[r]{-1}}
\put(252.0,468.0){\rule[-0.200pt]{4.818pt}{0.400pt}}
\put(252,613){\makebox(0,0)[r]{ 0}}
\put(252.0,613.0){\rule[-0.200pt]{4.818pt}{0.400pt}}
\put(252,758){\makebox(0,0)[r]{ 1}}
\put(252.0,758.0){\rule[-0.200pt]{4.818pt}{0.400pt}}
\put(99.0,593.0){\rule[-0.200pt]{0.400pt}{4.818pt}}
\put(99,552){\makebox(0,0){-0.5}}
\put(99.0,613.0){\rule[-0.200pt]{0.400pt}{4.818pt}}
\put(272.0,593.0){\rule[-0.200pt]{0.400pt}{4.818pt}}
\put(272,552){\makebox(0,0){ 0}}
\put(272.0,613.0){\rule[-0.200pt]{0.400pt}{4.818pt}}
\put(445.0,593.0){\rule[-0.200pt]{0.400pt}{4.818pt}}
\put(445,552){\makebox(0,0){ 0.5}}
\put(445.0,613.0){\rule[-0.200pt]{0.400pt}{4.818pt}}
\put(618.0,593.0){\rule[-0.200pt]{0.400pt}{4.818pt}}
\put(618,552){\makebox(0,0){ 1}}
\put(618.0,613.0){\rule[-0.200pt]{0.400pt}{4.818pt}}
\put(791.0,593.0){\rule[-0.200pt]{0.400pt}{4.818pt}}
\put(791,552){\makebox(0,0){ 1.5}}
\put(791.0,613.0){\rule[-0.200pt]{0.400pt}{4.818pt}}
\put(964.0,593.0){\rule[-0.200pt]{0.400pt}{4.818pt}}
\put(964,552){\makebox(0,0){ 2}}
\put(964.0,613.0){\rule[-0.200pt]{0.400pt}{4.818pt}}
\put(1138.0,593.0){\rule[-0.200pt]{0.400pt}{4.818pt}}
\put(1138,552){\makebox(0,0){ 2.5}}
\put(1138.0,613.0){\rule[-0.200pt]{0.400pt}{4.818pt}}
\put(1311.0,593.0){\rule[-0.200pt]{0.400pt}{4.818pt}}
\put(1311,552){\makebox(0,0){ 3}}
\put(1311.0,613.0){\rule[-0.200pt]{0.400pt}{4.818pt}}
\put(30.0,613.0){\rule[-0.200pt]{341.837pt}{0.400pt}}
\put(272.0,40.0){\rule[-0.200pt]{0.400pt}{197.297pt}}
\put(30.0,40.0){\rule[-0.200pt]{0.400pt}{197.297pt}}
\put(30.0,40.0){\rule[-0.200pt]{341.837pt}{0.400pt}}
\put(1449.0,40.0){\rule[-0.200pt]{0.400pt}{197.297pt}}
\put(30.0,859.0){\rule[-0.200pt]{341.837pt}{0.400pt}}
\put(1380,569){\makebox(0,0)[l]{$x$}}
\put(151,808){\makebox(0,0)[l]{$W(x)$}}
\put(145,468){\usebox{\plotpoint}}
\multiput(145.61,468.00)(0.447,6.714){3}{\rule{0.108pt}{4.233pt}}
\multiput(144.17,468.00)(3.000,22.214){2}{\rule{0.400pt}{2.117pt}}
\multiput(148.60,499.00)(0.468,1.651){5}{\rule{0.113pt}{1.300pt}}
\multiput(147.17,499.00)(4.000,9.302){2}{\rule{0.400pt}{0.650pt}}
\multiput(152.61,511.00)(0.447,1.579){3}{\rule{0.108pt}{1.167pt}}
\multiput(151.17,511.00)(3.000,5.579){2}{\rule{0.400pt}{0.583pt}}
\multiput(155.60,519.00)(0.468,1.066){5}{\rule{0.113pt}{0.900pt}}
\multiput(154.17,519.00)(4.000,6.132){2}{\rule{0.400pt}{0.450pt}}
\multiput(159.61,527.00)(0.447,1.132){3}{\rule{0.108pt}{0.900pt}}
\multiput(158.17,527.00)(3.000,4.132){2}{\rule{0.400pt}{0.450pt}}
\multiput(162.60,533.00)(0.468,0.627){5}{\rule{0.113pt}{0.600pt}}
\multiput(161.17,533.00)(4.000,3.755){2}{\rule{0.400pt}{0.300pt}}
\multiput(166.61,538.00)(0.447,0.909){3}{\rule{0.108pt}{0.767pt}}
\multiput(165.17,538.00)(3.000,3.409){2}{\rule{0.400pt}{0.383pt}}
\multiput(169.00,543.60)(0.481,0.468){5}{\rule{0.500pt}{0.113pt}}
\multiput(169.00,542.17)(2.962,4.000){2}{\rule{0.250pt}{0.400pt}}
\multiput(173.61,547.00)(0.447,0.685){3}{\rule{0.108pt}{0.633pt}}
\multiput(172.17,547.00)(3.000,2.685){2}{\rule{0.400pt}{0.317pt}}
\multiput(176.00,551.60)(0.481,0.468){5}{\rule{0.500pt}{0.113pt}}
\multiput(176.00,550.17)(2.962,4.000){2}{\rule{0.250pt}{0.400pt}}
\multiput(180.00,555.61)(0.462,0.447){3}{\rule{0.500pt}{0.108pt}}
\multiput(180.00,554.17)(1.962,3.000){2}{\rule{0.250pt}{0.400pt}}
\multiput(183.00,558.61)(0.462,0.447){3}{\rule{0.500pt}{0.108pt}}
\multiput(183.00,557.17)(1.962,3.000){2}{\rule{0.250pt}{0.400pt}}
\multiput(186.00,561.60)(0.481,0.468){5}{\rule{0.500pt}{0.113pt}}
\multiput(186.00,560.17)(2.962,4.000){2}{\rule{0.250pt}{0.400pt}}
\put(190,565.17){\rule{0.700pt}{0.400pt}}
\multiput(190.00,564.17)(1.547,2.000){2}{\rule{0.350pt}{0.400pt}}
\multiput(193.00,567.61)(0.685,0.447){3}{\rule{0.633pt}{0.108pt}}
\multiput(193.00,566.17)(2.685,3.000){2}{\rule{0.317pt}{0.400pt}}
\multiput(197.00,570.61)(0.462,0.447){3}{\rule{0.500pt}{0.108pt}}
\multiput(197.00,569.17)(1.962,3.000){2}{\rule{0.250pt}{0.400pt}}
\multiput(200.00,573.61)(0.685,0.447){3}{\rule{0.633pt}{0.108pt}}
\multiput(200.00,572.17)(2.685,3.000){2}{\rule{0.317pt}{0.400pt}}
\put(204,576.17){\rule{0.700pt}{0.400pt}}
\multiput(204.00,575.17)(1.547,2.000){2}{\rule{0.350pt}{0.400pt}}
\put(207,578.17){\rule{0.900pt}{0.400pt}}
\multiput(207.00,577.17)(2.132,2.000){2}{\rule{0.450pt}{0.400pt}}
\multiput(211.00,580.61)(0.462,0.447){3}{\rule{0.500pt}{0.108pt}}
\multiput(211.00,579.17)(1.962,3.000){2}{\rule{0.250pt}{0.400pt}}
\put(214,583.17){\rule{0.900pt}{0.400pt}}
\multiput(214.00,582.17)(2.132,2.000){2}{\rule{0.450pt}{0.400pt}}
\put(218,585.17){\rule{0.700pt}{0.400pt}}
\multiput(218.00,584.17)(1.547,2.000){2}{\rule{0.350pt}{0.400pt}}
\put(221,587.17){\rule{0.900pt}{0.400pt}}
\multiput(221.00,586.17)(2.132,2.000){2}{\rule{0.450pt}{0.400pt}}
\put(225,589.17){\rule{0.700pt}{0.400pt}}
\multiput(225.00,588.17)(1.547,2.000){2}{\rule{0.350pt}{0.400pt}}
\put(228,591.17){\rule{0.700pt}{0.400pt}}
\multiput(228.00,590.17)(1.547,2.000){2}{\rule{0.350pt}{0.400pt}}
\put(231,593.17){\rule{0.900pt}{0.400pt}}
\multiput(231.00,592.17)(2.132,2.000){2}{\rule{0.450pt}{0.400pt}}
\put(235,595.17){\rule{0.700pt}{0.400pt}}
\multiput(235.00,594.17)(1.547,2.000){2}{\rule{0.350pt}{0.400pt}}
\put(238,597.17){\rule{0.900pt}{0.400pt}}
\multiput(238.00,596.17)(2.132,2.000){2}{\rule{0.450pt}{0.400pt}}
\put(242,598.67){\rule{0.723pt}{0.400pt}}
\multiput(242.00,598.17)(1.500,1.000){2}{\rule{0.361pt}{0.400pt}}
\put(245,600.17){\rule{0.900pt}{0.400pt}}
\multiput(245.00,599.17)(2.132,2.000){2}{\rule{0.450pt}{0.400pt}}
\put(249,602.17){\rule{0.700pt}{0.400pt}}
\multiput(249.00,601.17)(1.547,2.000){2}{\rule{0.350pt}{0.400pt}}
\put(252,603.67){\rule{0.964pt}{0.400pt}}
\multiput(252.00,603.17)(2.000,1.000){2}{\rule{0.482pt}{0.400pt}}
\put(256,605.17){\rule{0.700pt}{0.400pt}}
\multiput(256.00,604.17)(1.547,2.000){2}{\rule{0.350pt}{0.400pt}}
\put(259,606.67){\rule{0.964pt}{0.400pt}}
\multiput(259.00,606.17)(2.000,1.000){2}{\rule{0.482pt}{0.400pt}}
\put(263,608.17){\rule{0.700pt}{0.400pt}}
\multiput(263.00,607.17)(1.547,2.000){2}{\rule{0.350pt}{0.400pt}}
\put(266,609.67){\rule{0.964pt}{0.400pt}}
\multiput(266.00,609.17)(2.000,1.000){2}{\rule{0.482pt}{0.400pt}}
\put(270,611.17){\rule{0.700pt}{0.400pt}}
\multiput(270.00,610.17)(1.547,2.000){2}{\rule{0.350pt}{0.400pt}}
\put(273,612.67){\rule{0.723pt}{0.400pt}}
\multiput(273.00,612.17)(1.500,1.000){2}{\rule{0.361pt}{0.400pt}}
\put(276,614.17){\rule{0.900pt}{0.400pt}}
\multiput(276.00,613.17)(2.132,2.000){2}{\rule{0.450pt}{0.400pt}}
\put(280,615.67){\rule{0.723pt}{0.400pt}}
\multiput(280.00,615.17)(1.500,1.000){2}{\rule{0.361pt}{0.400pt}}
\put(283,616.67){\rule{0.964pt}{0.400pt}}
\multiput(283.00,616.17)(2.000,1.000){2}{\rule{0.482pt}{0.400pt}}
\put(287,618.17){\rule{0.700pt}{0.400pt}}
\multiput(287.00,617.17)(1.547,2.000){2}{\rule{0.350pt}{0.400pt}}
\put(290,619.67){\rule{0.964pt}{0.400pt}}
\multiput(290.00,619.17)(2.000,1.000){2}{\rule{0.482pt}{0.400pt}}
\put(294,620.67){\rule{0.723pt}{0.400pt}}
\multiput(294.00,620.17)(1.500,1.000){2}{\rule{0.361pt}{0.400pt}}
\put(297,622.17){\rule{0.900pt}{0.400pt}}
\multiput(297.00,621.17)(2.132,2.000){2}{\rule{0.450pt}{0.400pt}}
\put(301,623.67){\rule{0.723pt}{0.400pt}}
\multiput(301.00,623.17)(1.500,1.000){2}{\rule{0.361pt}{0.400pt}}
\put(304,624.67){\rule{0.964pt}{0.400pt}}
\multiput(304.00,624.17)(2.000,1.000){2}{\rule{0.482pt}{0.400pt}}
\put(308,625.67){\rule{0.723pt}{0.400pt}}
\multiput(308.00,625.17)(1.500,1.000){2}{\rule{0.361pt}{0.400pt}}
\put(311,626.67){\rule{0.964pt}{0.400pt}}
\multiput(311.00,626.17)(2.000,1.000){2}{\rule{0.482pt}{0.400pt}}
\put(315,628.17){\rule{0.700pt}{0.400pt}}
\multiput(315.00,627.17)(1.547,2.000){2}{\rule{0.350pt}{0.400pt}}
\put(318,629.67){\rule{0.723pt}{0.400pt}}
\multiput(318.00,629.17)(1.500,1.000){2}{\rule{0.361pt}{0.400pt}}
\put(321,630.67){\rule{0.964pt}{0.400pt}}
\multiput(321.00,630.17)(2.000,1.000){2}{\rule{0.482pt}{0.400pt}}
\put(325,631.67){\rule{0.723pt}{0.400pt}}
\multiput(325.00,631.17)(1.500,1.000){2}{\rule{0.361pt}{0.400pt}}
\put(328,632.67){\rule{0.964pt}{0.400pt}}
\multiput(328.00,632.17)(2.000,1.000){2}{\rule{0.482pt}{0.400pt}}
\put(332,633.67){\rule{0.723pt}{0.400pt}}
\multiput(332.00,633.17)(1.500,1.000){2}{\rule{0.361pt}{0.400pt}}
\put(335,634.67){\rule{0.964pt}{0.400pt}}
\multiput(335.00,634.17)(2.000,1.000){2}{\rule{0.482pt}{0.400pt}}
\put(339,635.67){\rule{0.723pt}{0.400pt}}
\multiput(339.00,635.17)(1.500,1.000){2}{\rule{0.361pt}{0.400pt}}
\put(342,636.67){\rule{0.964pt}{0.400pt}}
\multiput(342.00,636.17)(2.000,1.000){2}{\rule{0.482pt}{0.400pt}}
\put(346,637.67){\rule{0.723pt}{0.400pt}}
\multiput(346.00,637.17)(1.500,1.000){2}{\rule{0.361pt}{0.400pt}}
\put(349,638.67){\rule{0.964pt}{0.400pt}}
\multiput(349.00,638.17)(2.000,1.000){2}{\rule{0.482pt}{0.400pt}}
\put(353,639.67){\rule{0.723pt}{0.400pt}}
\multiput(353.00,639.17)(1.500,1.000){2}{\rule{0.361pt}{0.400pt}}
\put(356,640.67){\rule{0.964pt}{0.400pt}}
\multiput(356.00,640.17)(2.000,1.000){2}{\rule{0.482pt}{0.400pt}}
\put(360,641.67){\rule{0.723pt}{0.400pt}}
\multiput(360.00,641.17)(1.500,1.000){2}{\rule{0.361pt}{0.400pt}}
\put(363,642.67){\rule{0.723pt}{0.400pt}}
\multiput(363.00,642.17)(1.500,1.000){2}{\rule{0.361pt}{0.400pt}}
\put(366,643.67){\rule{0.964pt}{0.400pt}}
\multiput(366.00,643.17)(2.000,1.000){2}{\rule{0.482pt}{0.400pt}}
\put(370,644.67){\rule{0.723pt}{0.400pt}}
\multiput(370.00,644.17)(1.500,1.000){2}{\rule{0.361pt}{0.400pt}}
\put(373,645.67){\rule{0.964pt}{0.400pt}}
\multiput(373.00,645.17)(2.000,1.000){2}{\rule{0.482pt}{0.400pt}}
\put(377,646.67){\rule{0.723pt}{0.400pt}}
\multiput(377.00,646.17)(1.500,1.000){2}{\rule{0.361pt}{0.400pt}}
\put(380,647.67){\rule{0.964pt}{0.400pt}}
\multiput(380.00,647.17)(2.000,1.000){2}{\rule{0.482pt}{0.400pt}}
\put(384,648.67){\rule{0.723pt}{0.400pt}}
\multiput(384.00,648.17)(1.500,1.000){2}{\rule{0.361pt}{0.400pt}}
\put(387,649.67){\rule{0.964pt}{0.400pt}}
\multiput(387.00,649.17)(2.000,1.000){2}{\rule{0.482pt}{0.400pt}}
\put(391,650.67){\rule{0.723pt}{0.400pt}}
\multiput(391.00,650.17)(1.500,1.000){2}{\rule{0.361pt}{0.400pt}}
\put(398,651.67){\rule{0.723pt}{0.400pt}}
\multiput(398.00,651.17)(1.500,1.000){2}{\rule{0.361pt}{0.400pt}}
\put(401,652.67){\rule{0.964pt}{0.400pt}}
\multiput(401.00,652.17)(2.000,1.000){2}{\rule{0.482pt}{0.400pt}}
\put(405,653.67){\rule{0.723pt}{0.400pt}}
\multiput(405.00,653.17)(1.500,1.000){2}{\rule{0.361pt}{0.400pt}}
\put(408,654.67){\rule{0.723pt}{0.400pt}}
\multiput(408.00,654.17)(1.500,1.000){2}{\rule{0.361pt}{0.400pt}}
\put(411,655.67){\rule{0.964pt}{0.400pt}}
\multiput(411.00,655.17)(2.000,1.000){2}{\rule{0.482pt}{0.400pt}}
\put(394.0,652.0){\rule[-0.200pt]{0.964pt}{0.400pt}}
\put(418,656.67){\rule{0.964pt}{0.400pt}}
\multiput(418.00,656.17)(2.000,1.000){2}{\rule{0.482pt}{0.400pt}}
\put(422,657.67){\rule{0.723pt}{0.400pt}}
\multiput(422.00,657.17)(1.500,1.000){2}{\rule{0.361pt}{0.400pt}}
\put(425,658.67){\rule{0.964pt}{0.400pt}}
\multiput(425.00,658.17)(2.000,1.000){2}{\rule{0.482pt}{0.400pt}}
\put(429,659.67){\rule{0.723pt}{0.400pt}}
\multiput(429.00,659.17)(1.500,1.000){2}{\rule{0.361pt}{0.400pt}}
\put(415.0,657.0){\rule[-0.200pt]{0.723pt}{0.400pt}}
\put(436,660.67){\rule{0.723pt}{0.400pt}}
\multiput(436.00,660.17)(1.500,1.000){2}{\rule{0.361pt}{0.400pt}}
\put(439,661.67){\rule{0.964pt}{0.400pt}}
\multiput(439.00,661.17)(2.000,1.000){2}{\rule{0.482pt}{0.400pt}}
\put(443,662.67){\rule{0.723pt}{0.400pt}}
\multiput(443.00,662.17)(1.500,1.000){2}{\rule{0.361pt}{0.400pt}}
\put(432.0,661.0){\rule[-0.200pt]{0.964pt}{0.400pt}}
\put(450,663.67){\rule{0.723pt}{0.400pt}}
\multiput(450.00,663.17)(1.500,1.000){2}{\rule{0.361pt}{0.400pt}}
\put(453,664.67){\rule{0.723pt}{0.400pt}}
\multiput(453.00,664.17)(1.500,1.000){2}{\rule{0.361pt}{0.400pt}}
\put(456,665.67){\rule{0.964pt}{0.400pt}}
\multiput(456.00,665.17)(2.000,1.000){2}{\rule{0.482pt}{0.400pt}}
\put(446.0,664.0){\rule[-0.200pt]{0.964pt}{0.400pt}}
\put(463,666.67){\rule{0.964pt}{0.400pt}}
\multiput(463.00,666.17)(2.000,1.000){2}{\rule{0.482pt}{0.400pt}}
\put(467,667.67){\rule{0.723pt}{0.400pt}}
\multiput(467.00,667.17)(1.500,1.000){2}{\rule{0.361pt}{0.400pt}}
\put(470,668.67){\rule{0.964pt}{0.400pt}}
\multiput(470.00,668.17)(2.000,1.000){2}{\rule{0.482pt}{0.400pt}}
\put(460.0,667.0){\rule[-0.200pt]{0.723pt}{0.400pt}}
\put(477,669.67){\rule{0.964pt}{0.400pt}}
\multiput(477.00,669.17)(2.000,1.000){2}{\rule{0.482pt}{0.400pt}}
\put(481,670.67){\rule{0.723pt}{0.400pt}}
\multiput(481.00,670.17)(1.500,1.000){2}{\rule{0.361pt}{0.400pt}}
\put(474.0,670.0){\rule[-0.200pt]{0.723pt}{0.400pt}}
\put(488,671.67){\rule{0.723pt}{0.400pt}}
\multiput(488.00,671.17)(1.500,1.000){2}{\rule{0.361pt}{0.400pt}}
\put(491,672.67){\rule{0.964pt}{0.400pt}}
\multiput(491.00,672.17)(2.000,1.000){2}{\rule{0.482pt}{0.400pt}}
\put(484.0,672.0){\rule[-0.200pt]{0.964pt}{0.400pt}}
\put(498,673.67){\rule{0.723pt}{0.400pt}}
\multiput(498.00,673.17)(1.500,1.000){2}{\rule{0.361pt}{0.400pt}}
\put(501,674.67){\rule{0.964pt}{0.400pt}}
\multiput(501.00,674.17)(2.000,1.000){2}{\rule{0.482pt}{0.400pt}}
\put(495.0,674.0){\rule[-0.200pt]{0.723pt}{0.400pt}}
\put(508,675.67){\rule{0.964pt}{0.400pt}}
\multiput(508.00,675.17)(2.000,1.000){2}{\rule{0.482pt}{0.400pt}}
\put(512,676.67){\rule{0.723pt}{0.400pt}}
\multiput(512.00,676.17)(1.500,1.000){2}{\rule{0.361pt}{0.400pt}}
\put(505.0,676.0){\rule[-0.200pt]{0.723pt}{0.400pt}}
\put(519,677.67){\rule{0.723pt}{0.400pt}}
\multiput(519.00,677.17)(1.500,1.000){2}{\rule{0.361pt}{0.400pt}}
\put(515.0,678.0){\rule[-0.200pt]{0.964pt}{0.400pt}}
\put(526,678.67){\rule{0.723pt}{0.400pt}}
\multiput(526.00,678.17)(1.500,1.000){2}{\rule{0.361pt}{0.400pt}}
\put(529,679.67){\rule{0.964pt}{0.400pt}}
\multiput(529.00,679.17)(2.000,1.000){2}{\rule{0.482pt}{0.400pt}}
\put(522.0,679.0){\rule[-0.200pt]{0.964pt}{0.400pt}}
\put(536,680.67){\rule{0.723pt}{0.400pt}}
\multiput(536.00,680.17)(1.500,1.000){2}{\rule{0.361pt}{0.400pt}}
\put(539,681.67){\rule{0.964pt}{0.400pt}}
\multiput(539.00,681.17)(2.000,1.000){2}{\rule{0.482pt}{0.400pt}}
\put(533.0,681.0){\rule[-0.200pt]{0.723pt}{0.400pt}}
\put(546,682.67){\rule{0.964pt}{0.400pt}}
\multiput(546.00,682.17)(2.000,1.000){2}{\rule{0.482pt}{0.400pt}}
\put(543.0,683.0){\rule[-0.200pt]{0.723pt}{0.400pt}}
\put(553,683.67){\rule{0.964pt}{0.400pt}}
\multiput(553.00,683.17)(2.000,1.000){2}{\rule{0.482pt}{0.400pt}}
\put(557,684.67){\rule{0.723pt}{0.400pt}}
\multiput(557.00,684.17)(1.500,1.000){2}{\rule{0.361pt}{0.400pt}}
\put(550.0,684.0){\rule[-0.200pt]{0.723pt}{0.400pt}}
\put(564,685.67){\rule{0.723pt}{0.400pt}}
\multiput(564.00,685.17)(1.500,1.000){2}{\rule{0.361pt}{0.400pt}}
\put(560.0,686.0){\rule[-0.200pt]{0.964pt}{0.400pt}}
\put(571,686.67){\rule{0.723pt}{0.400pt}}
\multiput(571.00,686.17)(1.500,1.000){2}{\rule{0.361pt}{0.400pt}}
\put(567.0,687.0){\rule[-0.200pt]{0.964pt}{0.400pt}}
\put(578,687.67){\rule{0.723pt}{0.400pt}}
\multiput(578.00,687.17)(1.500,1.000){2}{\rule{0.361pt}{0.400pt}}
\put(574.0,688.0){\rule[-0.200pt]{0.964pt}{0.400pt}}
\put(584,688.67){\rule{0.964pt}{0.400pt}}
\multiput(584.00,688.17)(2.000,1.000){2}{\rule{0.482pt}{0.400pt}}
\put(588,689.67){\rule{0.723pt}{0.400pt}}
\multiput(588.00,689.17)(1.500,1.000){2}{\rule{0.361pt}{0.400pt}}
\put(581.0,689.0){\rule[-0.200pt]{0.723pt}{0.400pt}}
\put(595,690.67){\rule{0.723pt}{0.400pt}}
\multiput(595.00,690.17)(1.500,1.000){2}{\rule{0.361pt}{0.400pt}}
\put(591.0,691.0){\rule[-0.200pt]{0.964pt}{0.400pt}}
\put(602,691.67){\rule{0.723pt}{0.400pt}}
\multiput(602.00,691.17)(1.500,1.000){2}{\rule{0.361pt}{0.400pt}}
\put(598.0,692.0){\rule[-0.200pt]{0.964pt}{0.400pt}}
\put(609,692.67){\rule{0.723pt}{0.400pt}}
\multiput(609.00,692.17)(1.500,1.000){2}{\rule{0.361pt}{0.400pt}}
\put(605.0,693.0){\rule[-0.200pt]{0.964pt}{0.400pt}}
\put(616,693.67){\rule{0.723pt}{0.400pt}}
\multiput(616.00,693.17)(1.500,1.000){2}{\rule{0.361pt}{0.400pt}}
\put(612.0,694.0){\rule[-0.200pt]{0.964pt}{0.400pt}}
\put(623,694.67){\rule{0.723pt}{0.400pt}}
\multiput(623.00,694.17)(1.500,1.000){2}{\rule{0.361pt}{0.400pt}}
\put(619.0,695.0){\rule[-0.200pt]{0.964pt}{0.400pt}}
\put(629,695.67){\rule{0.964pt}{0.400pt}}
\multiput(629.00,695.17)(2.000,1.000){2}{\rule{0.482pt}{0.400pt}}
\put(626.0,696.0){\rule[-0.200pt]{0.723pt}{0.400pt}}
\put(636,696.67){\rule{0.964pt}{0.400pt}}
\multiput(636.00,696.17)(2.000,1.000){2}{\rule{0.482pt}{0.400pt}}
\put(633.0,697.0){\rule[-0.200pt]{0.723pt}{0.400pt}}
\put(643,697.67){\rule{0.964pt}{0.400pt}}
\multiput(643.00,697.17)(2.000,1.000){2}{\rule{0.482pt}{0.400pt}}
\put(640.0,698.0){\rule[-0.200pt]{0.723pt}{0.400pt}}
\put(650,698.67){\rule{0.964pt}{0.400pt}}
\multiput(650.00,698.17)(2.000,1.000){2}{\rule{0.482pt}{0.400pt}}
\put(647.0,699.0){\rule[-0.200pt]{0.723pt}{0.400pt}}
\put(657,699.67){\rule{0.964pt}{0.400pt}}
\multiput(657.00,699.17)(2.000,1.000){2}{\rule{0.482pt}{0.400pt}}
\put(654.0,700.0){\rule[-0.200pt]{0.723pt}{0.400pt}}
\put(664,700.67){\rule{0.964pt}{0.400pt}}
\multiput(664.00,700.17)(2.000,1.000){2}{\rule{0.482pt}{0.400pt}}
\put(661.0,701.0){\rule[-0.200pt]{0.723pt}{0.400pt}}
\put(671,701.67){\rule{0.723pt}{0.400pt}}
\multiput(671.00,701.17)(1.500,1.000){2}{\rule{0.361pt}{0.400pt}}
\put(668.0,702.0){\rule[-0.200pt]{0.723pt}{0.400pt}}
\put(678,702.67){\rule{0.723pt}{0.400pt}}
\multiput(678.00,702.17)(1.500,1.000){2}{\rule{0.361pt}{0.400pt}}
\put(674.0,703.0){\rule[-0.200pt]{0.964pt}{0.400pt}}
\put(685,703.67){\rule{0.723pt}{0.400pt}}
\multiput(685.00,703.17)(1.500,1.000){2}{\rule{0.361pt}{0.400pt}}
\put(681.0,704.0){\rule[-0.200pt]{0.964pt}{0.400pt}}
\put(692,704.67){\rule{0.723pt}{0.400pt}}
\multiput(692.00,704.17)(1.500,1.000){2}{\rule{0.361pt}{0.400pt}}
\put(688.0,705.0){\rule[-0.200pt]{0.964pt}{0.400pt}}
\put(699,705.67){\rule{0.723pt}{0.400pt}}
\multiput(699.00,705.17)(1.500,1.000){2}{\rule{0.361pt}{0.400pt}}
\put(695.0,706.0){\rule[-0.200pt]{0.964pt}{0.400pt}}
\put(706,706.67){\rule{0.723pt}{0.400pt}}
\multiput(706.00,706.17)(1.500,1.000){2}{\rule{0.361pt}{0.400pt}}
\put(702.0,707.0){\rule[-0.200pt]{0.964pt}{0.400pt}}
\put(716,707.67){\rule{0.723pt}{0.400pt}}
\multiput(716.00,707.17)(1.500,1.000){2}{\rule{0.361pt}{0.400pt}}
\put(709.0,708.0){\rule[-0.200pt]{1.686pt}{0.400pt}}
\put(723,708.67){\rule{0.723pt}{0.400pt}}
\multiput(723.00,708.17)(1.500,1.000){2}{\rule{0.361pt}{0.400pt}}
\put(719.0,709.0){\rule[-0.200pt]{0.964pt}{0.400pt}}
\put(730,709.67){\rule{0.723pt}{0.400pt}}
\multiput(730.00,709.17)(1.500,1.000){2}{\rule{0.361pt}{0.400pt}}
\put(726.0,710.0){\rule[-0.200pt]{0.964pt}{0.400pt}}
\put(737,710.67){\rule{0.723pt}{0.400pt}}
\multiput(737.00,710.17)(1.500,1.000){2}{\rule{0.361pt}{0.400pt}}
\put(733.0,711.0){\rule[-0.200pt]{0.964pt}{0.400pt}}
\put(747,711.67){\rule{0.964pt}{0.400pt}}
\multiput(747.00,711.17)(2.000,1.000){2}{\rule{0.482pt}{0.400pt}}
\put(740.0,712.0){\rule[-0.200pt]{1.686pt}{0.400pt}}
\put(754,712.67){\rule{0.964pt}{0.400pt}}
\multiput(754.00,712.17)(2.000,1.000){2}{\rule{0.482pt}{0.400pt}}
\put(751.0,713.0){\rule[-0.200pt]{0.723pt}{0.400pt}}
\put(761,713.67){\rule{0.723pt}{0.400pt}}
\multiput(761.00,713.17)(1.500,1.000){2}{\rule{0.361pt}{0.400pt}}
\put(758.0,714.0){\rule[-0.200pt]{0.723pt}{0.400pt}}
\put(771,714.67){\rule{0.964pt}{0.400pt}}
\multiput(771.00,714.17)(2.000,1.000){2}{\rule{0.482pt}{0.400pt}}
\put(764.0,715.0){\rule[-0.200pt]{1.686pt}{0.400pt}}
\put(778,715.67){\rule{0.964pt}{0.400pt}}
\multiput(778.00,715.17)(2.000,1.000){2}{\rule{0.482pt}{0.400pt}}
\put(775.0,716.0){\rule[-0.200pt]{0.723pt}{0.400pt}}
\put(789,716.67){\rule{0.723pt}{0.400pt}}
\multiput(789.00,716.17)(1.500,1.000){2}{\rule{0.361pt}{0.400pt}}
\put(782.0,717.0){\rule[-0.200pt]{1.686pt}{0.400pt}}
\put(796,717.67){\rule{0.723pt}{0.400pt}}
\multiput(796.00,717.17)(1.500,1.000){2}{\rule{0.361pt}{0.400pt}}
\put(792.0,718.0){\rule[-0.200pt]{0.964pt}{0.400pt}}
\put(806,718.67){\rule{0.723pt}{0.400pt}}
\multiput(806.00,718.17)(1.500,1.000){2}{\rule{0.361pt}{0.400pt}}
\put(799.0,719.0){\rule[-0.200pt]{1.686pt}{0.400pt}}
\put(813,719.67){\rule{0.723pt}{0.400pt}}
\multiput(813.00,719.17)(1.500,1.000){2}{\rule{0.361pt}{0.400pt}}
\put(809.0,720.0){\rule[-0.200pt]{0.964pt}{0.400pt}}
\put(823,720.67){\rule{0.964pt}{0.400pt}}
\multiput(823.00,720.17)(2.000,1.000){2}{\rule{0.482pt}{0.400pt}}
\put(816.0,721.0){\rule[-0.200pt]{1.686pt}{0.400pt}}
\put(830,721.67){\rule{0.964pt}{0.400pt}}
\multiput(830.00,721.17)(2.000,1.000){2}{\rule{0.482pt}{0.400pt}}
\put(827.0,722.0){\rule[-0.200pt]{0.723pt}{0.400pt}}
\put(841,722.67){\rule{0.723pt}{0.400pt}}
\multiput(841.00,722.17)(1.500,1.000){2}{\rule{0.361pt}{0.400pt}}
\put(834.0,723.0){\rule[-0.200pt]{1.686pt}{0.400pt}}
\put(848,723.67){\rule{0.723pt}{0.400pt}}
\multiput(848.00,723.17)(1.500,1.000){2}{\rule{0.361pt}{0.400pt}}
\put(844.0,724.0){\rule[-0.200pt]{0.964pt}{0.400pt}}
\put(858,724.67){\rule{0.723pt}{0.400pt}}
\multiput(858.00,724.17)(1.500,1.000){2}{\rule{0.361pt}{0.400pt}}
\put(851.0,725.0){\rule[-0.200pt]{1.686pt}{0.400pt}}
\put(868,725.67){\rule{0.964pt}{0.400pt}}
\multiput(868.00,725.17)(2.000,1.000){2}{\rule{0.482pt}{0.400pt}}
\put(861.0,726.0){\rule[-0.200pt]{1.686pt}{0.400pt}}
\put(875,726.67){\rule{0.964pt}{0.400pt}}
\multiput(875.00,726.17)(2.000,1.000){2}{\rule{0.482pt}{0.400pt}}
\put(872.0,727.0){\rule[-0.200pt]{0.723pt}{0.400pt}}
\put(886,727.67){\rule{0.723pt}{0.400pt}}
\multiput(886.00,727.17)(1.500,1.000){2}{\rule{0.361pt}{0.400pt}}
\put(879.0,728.0){\rule[-0.200pt]{1.686pt}{0.400pt}}
\put(896,728.67){\rule{0.723pt}{0.400pt}}
\multiput(896.00,728.17)(1.500,1.000){2}{\rule{0.361pt}{0.400pt}}
\put(889.0,729.0){\rule[-0.200pt]{1.686pt}{0.400pt}}
\put(906,729.67){\rule{0.964pt}{0.400pt}}
\multiput(906.00,729.17)(2.000,1.000){2}{\rule{0.482pt}{0.400pt}}
\put(899.0,730.0){\rule[-0.200pt]{1.686pt}{0.400pt}}
\put(917,730.67){\rule{0.723pt}{0.400pt}}
\multiput(917.00,730.17)(1.500,1.000){2}{\rule{0.361pt}{0.400pt}}
\put(910.0,731.0){\rule[-0.200pt]{1.686pt}{0.400pt}}
\put(927,731.67){\rule{0.964pt}{0.400pt}}
\multiput(927.00,731.17)(2.000,1.000){2}{\rule{0.482pt}{0.400pt}}
\put(920.0,732.0){\rule[-0.200pt]{1.686pt}{0.400pt}}
\put(934,732.67){\rule{0.964pt}{0.400pt}}
\multiput(934.00,732.17)(2.000,1.000){2}{\rule{0.482pt}{0.400pt}}
\put(931.0,733.0){\rule[-0.200pt]{0.723pt}{0.400pt}}
\put(944,733.67){\rule{0.964pt}{0.400pt}}
\multiput(944.00,733.17)(2.000,1.000){2}{\rule{0.482pt}{0.400pt}}
\put(938.0,734.0){\rule[-0.200pt]{1.445pt}{0.400pt}}
\put(955,734.67){\rule{0.723pt}{0.400pt}}
\multiput(955.00,734.17)(1.500,1.000){2}{\rule{0.361pt}{0.400pt}}
\put(948.0,735.0){\rule[-0.200pt]{1.686pt}{0.400pt}}
\put(965,735.67){\rule{0.964pt}{0.400pt}}
\multiput(965.00,735.17)(2.000,1.000){2}{\rule{0.482pt}{0.400pt}}
\put(958.0,736.0){\rule[-0.200pt]{1.686pt}{0.400pt}}
\put(976,736.67){\rule{0.723pt}{0.400pt}}
\multiput(976.00,736.17)(1.500,1.000){2}{\rule{0.361pt}{0.400pt}}
\put(969.0,737.0){\rule[-0.200pt]{1.686pt}{0.400pt}}
\put(986,737.67){\rule{0.723pt}{0.400pt}}
\multiput(986.00,737.17)(1.500,1.000){2}{\rule{0.361pt}{0.400pt}}
\put(979.0,738.0){\rule[-0.200pt]{1.686pt}{0.400pt}}
\put(996,738.67){\rule{0.964pt}{0.400pt}}
\multiput(996.00,738.17)(2.000,1.000){2}{\rule{0.482pt}{0.400pt}}
\put(989.0,739.0){\rule[-0.200pt]{1.686pt}{0.400pt}}
\put(1010,739.67){\rule{0.964pt}{0.400pt}}
\multiput(1010.00,739.17)(2.000,1.000){2}{\rule{0.482pt}{0.400pt}}
\put(1000.0,740.0){\rule[-0.200pt]{2.409pt}{0.400pt}}
\put(1021,740.67){\rule{0.723pt}{0.400pt}}
\multiput(1021.00,740.17)(1.500,1.000){2}{\rule{0.361pt}{0.400pt}}
\put(1014.0,741.0){\rule[-0.200pt]{1.686pt}{0.400pt}}
\put(1031,741.67){\rule{0.723pt}{0.400pt}}
\multiput(1031.00,741.17)(1.500,1.000){2}{\rule{0.361pt}{0.400pt}}
\put(1024.0,742.0){\rule[-0.200pt]{1.686pt}{0.400pt}}
\put(1041,742.67){\rule{0.964pt}{0.400pt}}
\multiput(1041.00,742.17)(2.000,1.000){2}{\rule{0.482pt}{0.400pt}}
\put(1034.0,743.0){\rule[-0.200pt]{1.686pt}{0.400pt}}
\put(1052,743.67){\rule{0.723pt}{0.400pt}}
\multiput(1052.00,743.17)(1.500,1.000){2}{\rule{0.361pt}{0.400pt}}
\put(1045.0,744.0){\rule[-0.200pt]{1.686pt}{0.400pt}}
\put(1066,744.67){\rule{0.723pt}{0.400pt}}
\multiput(1066.00,744.17)(1.500,1.000){2}{\rule{0.361pt}{0.400pt}}
\put(1055.0,745.0){\rule[-0.200pt]{2.650pt}{0.400pt}}
\put(1076,745.67){\rule{0.723pt}{0.400pt}}
\multiput(1076.00,745.17)(1.500,1.000){2}{\rule{0.361pt}{0.400pt}}
\put(1069.0,746.0){\rule[-0.200pt]{1.686pt}{0.400pt}}
\put(1086,746.67){\rule{0.964pt}{0.400pt}}
\multiput(1086.00,746.17)(2.000,1.000){2}{\rule{0.482pt}{0.400pt}}
\put(1079.0,747.0){\rule[-0.200pt]{1.686pt}{0.400pt}}
\put(1100,747.67){\rule{0.964pt}{0.400pt}}
\multiput(1100.00,747.17)(2.000,1.000){2}{\rule{0.482pt}{0.400pt}}
\put(1090.0,748.0){\rule[-0.200pt]{2.409pt}{0.400pt}}
\put(1111,748.67){\rule{0.723pt}{0.400pt}}
\multiput(1111.00,748.17)(1.500,1.000){2}{\rule{0.361pt}{0.400pt}}
\put(1104.0,749.0){\rule[-0.200pt]{1.686pt}{0.400pt}}
\put(1124,749.67){\rule{0.964pt}{0.400pt}}
\multiput(1124.00,749.17)(2.000,1.000){2}{\rule{0.482pt}{0.400pt}}
\put(1114.0,750.0){\rule[-0.200pt]{2.409pt}{0.400pt}}
\put(1135,750.67){\rule{0.723pt}{0.400pt}}
\multiput(1135.00,750.17)(1.500,1.000){2}{\rule{0.361pt}{0.400pt}}
\put(1128.0,751.0){\rule[-0.200pt]{1.686pt}{0.400pt}}
\put(1149,751.67){\rule{0.723pt}{0.400pt}}
\multiput(1149.00,751.17)(1.500,1.000){2}{\rule{0.361pt}{0.400pt}}
\put(1138.0,752.0){\rule[-0.200pt]{2.650pt}{0.400pt}}
\put(1159,752.67){\rule{0.723pt}{0.400pt}}
\multiput(1159.00,752.17)(1.500,1.000){2}{\rule{0.361pt}{0.400pt}}
\put(1152.0,753.0){\rule[-0.200pt]{1.686pt}{0.400pt}}
\put(1173,753.67){\rule{0.723pt}{0.400pt}}
\multiput(1173.00,753.17)(1.500,1.000){2}{\rule{0.361pt}{0.400pt}}
\put(1162.0,754.0){\rule[-0.200pt]{2.650pt}{0.400pt}}
\put(1187,754.67){\rule{0.723pt}{0.400pt}}
\multiput(1187.00,754.17)(1.500,1.000){2}{\rule{0.361pt}{0.400pt}}
\put(1176.0,755.0){\rule[-0.200pt]{2.650pt}{0.400pt}}
\put(1197,755.67){\rule{0.964pt}{0.400pt}}
\multiput(1197.00,755.17)(2.000,1.000){2}{\rule{0.482pt}{0.400pt}}
\put(1190.0,756.0){\rule[-0.200pt]{1.686pt}{0.400pt}}
\put(1211,756.67){\rule{0.723pt}{0.400pt}}
\multiput(1211.00,756.17)(1.500,1.000){2}{\rule{0.361pt}{0.400pt}}
\put(1201.0,757.0){\rule[-0.200pt]{2.409pt}{0.400pt}}
\put(1225,757.67){\rule{0.723pt}{0.400pt}}
\multiput(1225.00,757.17)(1.500,1.000){2}{\rule{0.361pt}{0.400pt}}
\put(1214.0,758.0){\rule[-0.200pt]{2.650pt}{0.400pt}}
\put(1239,758.67){\rule{0.723pt}{0.400pt}}
\multiput(1239.00,758.17)(1.500,1.000){2}{\rule{0.361pt}{0.400pt}}
\put(1228.0,759.0){\rule[-0.200pt]{2.650pt}{0.400pt}}
\put(1249,759.67){\rule{0.723pt}{0.400pt}}
\multiput(1249.00,759.17)(1.500,1.000){2}{\rule{0.361pt}{0.400pt}}
\put(1242.0,760.0){\rule[-0.200pt]{1.686pt}{0.400pt}}
\put(1263,760.67){\rule{0.723pt}{0.400pt}}
\multiput(1263.00,760.17)(1.500,1.000){2}{\rule{0.361pt}{0.400pt}}
\put(1252.0,761.0){\rule[-0.200pt]{2.650pt}{0.400pt}}
\put(1277,761.67){\rule{0.723pt}{0.400pt}}
\multiput(1277.00,761.17)(1.500,1.000){2}{\rule{0.361pt}{0.400pt}}
\put(1266.0,762.0){\rule[-0.200pt]{2.650pt}{0.400pt}}
\put(1291,762.67){\rule{0.723pt}{0.400pt}}
\multiput(1291.00,762.17)(1.500,1.000){2}{\rule{0.361pt}{0.400pt}}
\put(1280.0,763.0){\rule[-0.200pt]{2.650pt}{0.400pt}}
\put(1304,763.67){\rule{0.964pt}{0.400pt}}
\multiput(1304.00,763.17)(2.000,1.000){2}{\rule{0.482pt}{0.400pt}}
\put(1294.0,764.0){\rule[-0.200pt]{2.409pt}{0.400pt}}
\put(1318,764.67){\rule{0.964pt}{0.400pt}}
\multiput(1318.00,764.17)(2.000,1.000){2}{\rule{0.482pt}{0.400pt}}
\put(1308.0,765.0){\rule[-0.200pt]{2.409pt}{0.400pt}}
\put(1332,765.67){\rule{0.964pt}{0.400pt}}
\multiput(1332.00,765.17)(2.000,1.000){2}{\rule{0.482pt}{0.400pt}}
\put(1322.0,766.0){\rule[-0.200pt]{2.409pt}{0.400pt}}
\put(1346,766.67){\rule{0.723pt}{0.400pt}}
\multiput(1346.00,766.17)(1.500,1.000){2}{\rule{0.361pt}{0.400pt}}
\put(1336.0,767.0){\rule[-0.200pt]{2.409pt}{0.400pt}}
\put(1363,767.67){\rule{0.964pt}{0.400pt}}
\multiput(1363.00,767.17)(2.000,1.000){2}{\rule{0.482pt}{0.400pt}}
\put(1349.0,768.0){\rule[-0.200pt]{3.373pt}{0.400pt}}
\put(1377,768.67){\rule{0.964pt}{0.400pt}}
\multiput(1377.00,768.17)(2.000,1.000){2}{\rule{0.482pt}{0.400pt}}
\put(1367.0,769.0){\rule[-0.200pt]{2.409pt}{0.400pt}}
\put(1391,769.67){\rule{0.723pt}{0.400pt}}
\multiput(1391.00,769.17)(1.500,1.000){2}{\rule{0.361pt}{0.400pt}}
\put(1381.0,770.0){\rule[-0.200pt]{2.409pt}{0.400pt}}
\put(1405,770.67){\rule{0.723pt}{0.400pt}}
\multiput(1405.00,770.17)(1.500,1.000){2}{\rule{0.361pt}{0.400pt}}
\put(1394.0,771.0){\rule[-0.200pt]{2.650pt}{0.400pt}}
\put(1422,771.67){\rule{0.964pt}{0.400pt}}
\multiput(1422.00,771.17)(2.000,1.000){2}{\rule{0.482pt}{0.400pt}}
\put(1408.0,772.0){\rule[-0.200pt]{3.373pt}{0.400pt}}
\put(1436,772.67){\rule{0.723pt}{0.400pt}}
\multiput(1436.00,772.17)(1.500,1.000){2}{\rule{0.361pt}{0.400pt}}
\put(1426.0,773.0){\rule[-0.200pt]{2.409pt}{0.400pt}}
\put(1439.0,774.0){\rule[-0.200pt]{2.409pt}{0.400pt}}
\put(145,468){\usebox{\plotpoint}}
\multiput(145,468)(1.677,-20.688){2}{\usebox{\plotpoint}}
\put(149.01,426.73){\usebox{\plotpoint}}
\put(153.51,406.47){\usebox{\plotpoint}}
\put(159.12,386.52){\usebox{\plotpoint}}
\put(164.81,366.58){\usebox{\plotpoint}}
\put(170.91,346.76){\usebox{\plotpoint}}
\put(177.20,327.00){\usebox{\plotpoint}}
\put(183.48,307.25){\usebox{\plotpoint}}
\put(189.82,287.51){\usebox{\plotpoint}}
\put(196.01,267.71){\usebox{\plotpoint}}
\put(201.74,247.78){\usebox{\plotpoint}}
\put(207.39,227.83){\usebox{\plotpoint}}
\put(213.10,207.90){\usebox{\plotpoint}}
\put(218.46,187.85){\usebox{\plotpoint}}
\put(223.09,167.63){\usebox{\plotpoint}}
\put(227.53,147.35){\usebox{\plotpoint}}
\put(231.22,126.93){\usebox{\plotpoint}}
\put(235.38,106.60){\usebox{\plotpoint}}
\put(238.71,86.11){\usebox{\plotpoint}}
\multiput(242,68)(2.684,-20.581){2}{\usebox{\plotpoint}}
\put(246,40){\usebox{\plotpoint}}
\put(30.0,40.0){\rule[-0.200pt]{0.400pt}{197.297pt}}
\put(30.0,40.0){\rule[-0.200pt]{341.837pt}{0.400pt}}
\put(1449.0,40.0){\rule[-0.200pt]{0.400pt}{197.297pt}}
\put(30.0,859.0){\rule[-0.200pt]{341.837pt}{0.400pt}}
\end{picture}